\pdfoutput=1

\documentclass[11pt,letterpaper]{article}
\usepackage{jheppub}
\usepackage[all]{xy}
\usepackage{xcolor}
\usepackage{url}
\usepackage{diagrams}
\def\ket#1{\langle #1 \rangle}

\DeclareMathOperator{\B}{B}
\DeclareMathOperator{\Conf}{Conf}
\DeclareMathOperator{\Gr}{Gr}
\DeclareMathOperator{\Li}{Li}

\preprint{Brown-HET-1654}

\title{Cluster Polylogarithms for Scattering Amplitudes}

\author[a]{John Golden,}
\author[a]{Miguel F.~Paulos,}
\author[a]{Marcus Spradlin,}
\author[a]{and Anastasia Volovich}

\affiliation[a]{Department of Physics, Brown University,\\
Box 1843,\\
Providence, RI 02912-1843,
USA}

\abstract{
Motivated by the cluster structure of two-loop scattering
amplitudes in $\mathcal{N}=4$ Yang-Mills theory we define
\emph{cluster polylogarithm functions}.  We find that all such
functions of weight 4 are made up of a single simple building
block associated to the $A_2$ cluster algebra. Adding the
requirement of locality on generalized Stasheff polytopes,
we find that these $A_2$ building blocks arrange themselves to
form a unique function associated to the $A_3$ cluster algebra.
This $A_3$ function manifests all of the cluster algebraic
structure
of the two-loop $n$-particle MHV amplitudes
for all $n$, and we use it to provide an explicit
representation for the most complicated part of the $n=7$ amplitude
as an example.
}

\begin{document}
\maketitle

\section{Introduction}
\label{sec:1}

There exists a vast mathematical literature on cluster algebras (see for
example~\cite{ClusterPortal}), and
also a large literature on the mathematical
structure of generalized polylogarithm functions.
One of the general themes to emerge from~\cite{Golden:2013xva}
was the observation
that the perturbative scattering
amplitudes in $\mathcal{N}=4$ supersymmetric
Yang-Mills (SYM) theory, which have also been the object of intense
study in recent years, tie these two topics intimately
together\footnote{See also~\cite{ArkaniHamed:2012nw} for a different
relation between cluster algebras and scattering amplitudes.}.

In this paper we take a few steps towards a systematic investigation of
the intersection between these fields of mathematics.
To that end we define and study
the simplest examples of
\emph{cluster polylogarithm functions}---pure transcendental
functions which ``depend on'' (in a way to be made precise below)
only the cluster coordinates of some cluster algebra.  Even the
mere existence of non-trivial examples of such functions is
not a priori obvious---for example, we will see that the Gr(3,5) cluster
algebra admits only a single non-trivial cluster function
of weight four.
Special functions of this kind are
apparently not known in the mathematical literature, but we know
they exist and
are likely to have remarkable properties since SYM theory
evidently
provides (in addition to numerous other generous mathematical gifts) at
least one infinite class of such functions: the two-loop
$n$-particle MHV amplitudes~\cite{futureWork}.

In addition to the purely mathematical motivation for exploring this new
class of special functions, our work also has a practical
application for physicists.
The symbol of all two-loop $n$-particle MHV amplitudes has been known
for almost three years from the work of Caron-Huot~\cite{CaronHuot:2011ky},
but it remains an interesting outstanding problem to write explicit analytic
formulas for these amplitudes\footnote{Considerable analytic progress
has been made both at two loops and (in some cases) far beyond
for special kinematic configurations, including for example
multi-Regge kinematics~\cite{Prygarin:2011gd,Bartels:2013jna,Bartels:2011xy,Dixon:2012yy,Pennington:2012zj},
the near-collinear limit~\cite{Basso:2013aha,Basso:2013vsa,Papathanasiou:2013uoa}, and
2d kinematics~\cite{Heslop:2010kq,Heslop:2011hv,Goddard:2012cx,Caron-Huot:2013vda}
(see also~\cite{Torres:2013vba} for comments on cluster structure in 2d kinematics).
Fully analytic
formulas for the \emph{differential} of the two-loop MHV amplitude for all $n$
were computed in~\cite{Golden:2013lha}.}.
So far this has been
accomplished~\cite{DelDuca:2009au,DelDuca:2010zg} only
for the very special case of $n=6$, where the result
surprisingly can be written entirely in terms of the classical
polylogarithm functions $\Li_k$~\cite{Goncharov:2010jf} (a curiosity which
we ``explain'' below).
To write analytic results for generic scattering amplitudes, even in SYM
theory, requires giving up the crutch of working with only the
relatively tame classical functions and entering the much
larger and wilder world of generalized polylogarithm functions.
Several impressive analytic results of this type have been
achieved
for higher-loop or non-MHV $n=6$ amplitudes in SYM
theory by Dixon and collaborators~\cite{Dixon:2011nj,Dixon:2013eka}.
Applying
similar technology at higher $n$ looks challenging because
the required computer power grows rapidly with $n$.
Ultimately this is due to the fact that absent other guidance,
one would run the risk of being forced
to work with a vastly overcomplete basis
of functions not specifically tailored to the problem at hand.

When studying $n>6$ amplitudes in SYM theory it is
desirable, for both practical as well as aesthetic reasons,
to seek out functional representations which
manifest (to the extent possible) all of the important
properties of an amplitude.
For example, the GSVV formula~\cite{Goncharov:2010jf},
unlike the previously known DDS formula, makes three important
properties of the two-loop $n=6$ MHV amplitude trivially manifest:  it is
classical, dihedral invariant, and real-valued everywhere inside the Euclidean
domain.  (There is also one interesting property which the GSVV
formula does not make manifest: the fact that it is positive-valued
everywhere inside the positive domain.)
Working with functional representatives which manifest as many properties
as possible has enormous practical advantages over working with a generic
basis of functions, and also helps to elucidate the
deeper mathematical structure of
the amplitudes.
Of course as time passes we may discover new, previously unnoticed
properties, allowing us the opportunity to further upgrade the class of
functions we work with.

Suppose we were to commission some
very special custom non-classical polylogarithm
function (or collection of functions) specifically suited for
the purpose of
expressing the two-loop $n$-point MHV amplitudes.
Based on what we know about these amplitudes today,
what properties should
we demand these special functions manifest?
The most basic property we should impose is that the symbol alphabet
should consist of the cluster $\mathcal{A}$-coordinates of
the
$\Gr(4,n)$ cluster algebra, a property of the amplitudes which is
manifest in the result of~\cite{CaronHuot:2011ky}.
We call such functions ``cluster $\mathcal{A}$-functions''.
For the special case of the $\Gr(4,6)$ algebra, relevant to
$n=6$ particle amplitudes,
functions of this type (and satisfying also various other physical
constraints) were extensively studied, and fully classified through
weight 6 in~\cite{Dixon:2011pw,Dixon:2011nj,Dixon:2013eka}.

Taking inspiration from this program of classifying allowed functions,
we consider here two additional properties
of the two-loop MHV amplitudes
which were recently observed
in~\cite{Golden:2013xva,futureWork,SimonsTalk}:  (1) the coproduct of these
amplitudes can be expressed entirely in terms of
$\mathcal{X}$-coordinates~\cite{FG03b}
on the cluster Poisson variety
$\Conf_n(\mathbb{P}^3)$ (or equivalently, the $\Gr(4,n)$ cluster algebra),
and moreover (2) that the $\Lambda^2 \B_2$ component of the coproduct
can be expressed entirely in terms of pairs of variables which
Poisson commute.

Remarkably we find that the two simplest non-trivial cluster polylogarithm
functions exactly fit the bill.  Specifically, we find
at transcendentality weight four
that the $\Gr(3,5)$ cluster algebra (also called $A_2$)
admits a unique non-trivial function
satisfying property (1), and the $\Gr(4,6)$ (or $A_3$) cluster algebra
admits a unique non-trivial function (itself a linear combination
of $A_2$ functions) which in addition satisfies property (2).
We have checked explicitly for a small handful of cluster algebras
that the functions associated with all
$A_2$ and $A_3$ subalgebras provide a complete basis for \emph{all}
weight-four functions satisfying these properties.
It is certainly an interesting mathematical problem to explore the
universe of cluster functions for general algebras, but
for the more limited purpose of expressing two-loop MHV amplitudes it seems
that the six-particle $A_3$ function is all we need.
Although the
structure of the $n$-point amplitude stabilizes at relatively small $n$
(that means that higher $n$ amplitudes can be written in terms of
building blocks
involving smaller values of $n$),
it is rather surprising that the basic two-loop
building block seems to involve only six particles.

In section~\ref{sec:2} we briefly review some mathematical background necessary
for formulating our definition of cluster functions.  In sections~\ref{sec:3}
and~\ref{sec:4}
respectively
we discuss the $A_2$ and $A_3$ functions, and in section~\ref{sec:5}
we comment on
the problem of expressing two-loop $n$-point MHV amplitudes in terms of
$A_3$ functions, providing an explicit result for $n=7$ as an example.

\section{Cluster polylogarithm functions}
\label{sec:2}

\subsection{Polylogarithm functions, symbols, and the coproduct}

We begin by recalling some elementary mathematical facts about polylogarithm
functions from~\cite{G91b,Goncharov-Galois2}
(see~\cite{Duhr:2011zq,Duhr:2012fh,Golden:2013xva,VerguTalk} for recent reviews
written for physicists).  To each such transcendental function of
weight $k$ is associated an element of the $k$-fold tensor
product of the multiplicative group of
rational functions modulo constants called its \emph{symbol}.
For example, the classical polylogarithm function $\Li_k(x)$ has
symbol
\begin{equation}
- (1 - x) \otimes
\underbrace{x \otimes \cdots \otimes x}_{k-1~{\rm times}}.
\end{equation}

A trivial way to make a function of weight $k$ is to multiply two
functions of lower weights $k_1,k_2$ with $k=k_1+k_2$.  It is often
useful to exclude such products from consideration and to focus on the
most complicated, intrinsically weight $k$, part of a function.
This may be accomplished via a projection operator $\rho$ which
annihilates all products of functions of lower weight.  It is defined
recursively by
\begin{equation}
\rho(a_1 \otimes \cdots \otimes a_k) =
\frac{k-1}{k} \left[
\rho(a_1 \otimes \cdots \otimes a_{k-1}) \otimes a_k
- \rho(a_2 \otimes \cdots \otimes a_k) \otimes a_1
\right]
\end{equation}
beginning with $\rho(a_1) = a_1$.
Here, in a slight abuse of notation which we will perpetuate throughout this section,
we display for simplicity not how $\rho$ acts on a general weight-$k$
function but rather how it acts on the symbol of such a function.

We use $\mathcal{L}_\bullet$ to denote the algebra of
polylogarithm functions
modulo products of functions of lower weight.
It is a commutative graded Hopf algebra with a coproduct
$\delta: \mathcal{L}_\bullet \mapsto \Lambda^2 \mathcal{L}_\bullet$
which satisfies $\delta^2 = 0$, giving it
the structure of a Lie coalgebra.
Explicitly, $\delta$ may be computed (again, at the level of symbols) by
\begin{equation}
\label{eq:deltadef}
\delta(a_1 \otimes \cdots \otimes a_k)
= \sum_{n=1}^{k-1}
(a_1 \otimes \cdots \otimes a_n) \bigwedge (a_{n+1} \otimes \cdots
\otimes a_k).
\end{equation}

We let $\B_k$ denote the Bloch group~\cite{Bl,Su} defined as the quotient of $\mathcal{L}_k$ by the
subspace of functional equations for the classical logarithm function $\Li_k$.
The case $k=1$ is trivial (any linear combination of logarithm functions
can be combined into a single logarithm)
so we simply write ``$x$'' to denote the function $\log x$ and therefore
denote $\mathcal{L}_1
= \mathbb{C}^*$, the multiplicative group of nonzero complex numbers.
For $k>1$ elements of $\B_k$ are finite
linear combinations of objects denoted by $\{x\}_k$, which can be
read as shorthand for the function $-\Li_k(-x)$.
These satisfy\footnote{The top line is an element of $\Lambda^2 \mathcal{L}_1$,
while the bottom is the element of the summand in $\Lambda^2 ( \mathcal{L}_{k-1} \oplus \mathcal{L}_1)$
given by vectors of the form $f_{k-1} \otimes f_1 - f_1 \otimes f_{k-1}$, and we use the standard
notation of denoting such an element simply by
$f_{k-1} \otimes f_1 \in \mathcal{L}_{k-1} \otimes \mathcal{L}_1$.}
\begin{equation}
\delta \{x\}_k = \begin{cases}
(1 + x) \bigwedge x & k = 2,\\
\{x\}_{k-1} \bigotimes x & k > 2.
\end{cases}
\end{equation}
For $k=2,3$ it is a theorem that $\mathcal{L}_k = \B_k$.
At weight 4, for the first time, the coproduct has two separate
components\footnote{These expressions are simply transcriptions of the components of eq.~(\ref{eq:deltadef}).
Specifically, eq.~(\ref{eq:twofive}) is the $n=2$ term in eq.~(\ref{eq:deltadef}), while
eq.~(\ref{eq:twosix}) is the sum of the $n=1$ and $n=3$ terms in eq.~(\ref{eq:deltadef}), with the
$n=1$ term receiving a minus sign when put into the order shown in eq.~(\ref{eq:twosix}) as implied
by the $\bigwedge$ in eq.~(\ref{eq:deltadef}).  We have
chosen to write some $\rho$'s explicitly in eqs.~(\ref{eq:twofive}) and~(\ref{eq:twosix}) but they
are not necessary since $\rho(x)$ and $x$ represent the same element in $\mathcal{L}_k$.  For the same reason,
we could have acted on both sides of the $\bigwedge$ in eq.~(\ref{eq:deltadef}) with the $\rho$ operator
without changing the meaning of that definition.}
\begin{align}
\label{eq:twofive}
\delta(a_1 \otimes a_2 \otimes a_3 \otimes a_4)\rvert_{\Lambda^2 \B_2}
&= \rho(a_1 \otimes a_2) \bigwedge \rho(a_3 \otimes a_4),\\
\delta(a_1 \otimes a_2 \otimes a_3 \otimes a_4)\rvert_{\B_3 \otimes \mathbb{C}^*} &= \rho(a_1 \otimes a_2 \otimes a_3) \bigotimes a_4 -
\rho(a_2 \otimes a_3 \otimes a_4) \bigotimes a_1.
\label{eq:twosix}
\end{align}
The classical function $\Li_4(x)$ has coproduct components
\begin{align}
        \delta \Li_4(x)\rvert_{\Lambda^2 \B_2}&=0,\\
        \delta \Li_4(x)\rvert_{\B_3 \otimes \mathbb{C}^*}&=-\{-x\}_3 \otimes x,
\end{align}
so it is clear that any polylogarithm function of weight 4 which has
a nonzero $\Lambda^2 \B_2$ content cannot possibly be written in terms
of classical functions.  It is moreover conjectured that the converse is true~\cite{G91b}\footnote{More generally,
it is conjectured that a weight-$k$ function $f_k$ can be written in terms of the classical
polylogarithm $\Li_k$ if and only if all components of $\delta f_k$ vanish except
possibly $\mathcal{L}_{k-1} \otimes \mathbb{C}^*$.}.
In this sense
we can say that it is the $\Lambda^2 \B_2$ coproduct component
which measures the ``non-trivial part'' of a weight-four polylogarithm function.

\subsection{Integrability}

Next we discuss the integrability condition which plays the
crucial role in the following two sections.
A second application of $\delta$ at weight 4
maps each of the two components to
$\B_2 \otimes \Lambda^2 \mathbb{C}^*$, as indicated in the diagram\footnote{Note that this is not
a commutative diagram; indeed according to eq~(\ref{eq:integrability}) it is better thought of as
an {\it anti}commutative diagram.}:
\begin{diagram}
&&\mathcal{L}_4\\
&\ldTo&&\rdTo\\
\Lambda^2 \B_2
&&&& \B_3 \otimes \mathbb{C}^*~, \\
&\rdTo&&\ldTo\\
&&\B_2 \otimes \Lambda^2 \mathbb{C}^*
\end{diagram}
where the bottom two arrows are given explicitly by
\begin{align}
\delta ( \{x\}_2 \wedge \{y \}_2 ) &=
\{y\}_2 \otimes (1 + x) \wedge x -
\{x\}_2 \otimes (1 + y) \wedge y,\\
\delta ( \{x\}_3 \otimes y )&=
\{x\}_2 \otimes x \wedge y.
\end{align}

Given arbitrary elements $b_{22} \in \Lambda^2 \B_2$
and $b_{31} \in \B_3 \otimes \mathbb{C}^*$, there does not necessarily
exist any function $f_4 \in \mathcal{L}_4$ whose coproduct
components are $b_{22}$ and
$b_{31}$.  A necessary and sufficient condition for such a function to
exist is that the integrability condition
\begin{equation}
\label{eq:integrability}
0 = \delta^2 f_4 =
\delta(b_{22}) + \delta(b_{31})
\end{equation}
is satisfied.
Equivalently, we can say that \emph{a pair $b_{22}, b_{31}$
satisfying~(\ref{eq:integrability}) uniquely determines a
weight-four polylogarithm function
(modulo products of functions of lower weight)}.

It is important to note that given any element $b_{22} \in \Lambda^2 \B_2$
there does exist some function $f_4$ with $b_{22}$ as its coproduct
component (indeed Goncharov has written down~\cite{Goncharov-Galois1} an explicit
map $\kappa: \Lambda^2 \B_2 \to \B_3  \otimes \mathbb{C}^*$ such that
the pair $b_{22}, \kappa(b_{22})$ satisfies~(\ref{eq:integrability})
for any $b_{22} \in \Lambda^2 \B_2$),
but for generic $b_{22}$ the $\B_3 \otimes \mathbb{C}^*$
component $\kappa(b_{22})$ of
that function will not have any cluster algebra structure
of the type we study below.

\subsection{Cluster $\mathcal{A}$- and $\mathcal{X}$-coordinates}

Next we provide a lightning review
(see~\cite{Golden:2013xva} for details)
of the types of variables
which make an appearance in the study of scattering amplitudes in SYM
theory:  cluster $\mathcal{A}$- and cluster $\mathcal{X}$-coordinates.
Much of what we have to say about cluster polylogarithm functions may
be interesting to investigate in the context of general algebras,
but we restrict our attention here largely to Grassmannian
cluster algebras, and in particular to the $\Gr(4,n)$
algebra relevant to the kinematic configuration space $\Conf_n(\mathbb{P}^3)$
of $n$-particle
scattering in SYM theory.

Examples of $\mathcal{A}$-coordinates on $\Gr(4,n)$
include the ordinary Pl\"ucker coordinates
$\ket{ijkl}$ as well as certain particular homogeneous polynomials in them
such as
\begin{equation}
\label{eq:aexample}
\begin{aligned}
\ket{a(bc)(de)(fg)} &\equiv \ket{abde}\ket{acfg} - \ket{abfg} \ket{acde},\\
\ket{ab(cde) \cap (fgh)} &\equiv \ket{acde}\ket{bfgh} -
\ket{bcde}\ket{afgh},
\end{aligned}
\end{equation}
while the $\mathcal{X}$-coordinates are certain cross-ratios which can be
built from
$\mathcal{A}$-coordinates.

For $n>7$ there exist arbitrarily more complicated
$\mathcal{A}$-coordinates on $\Gr(4,n)$.
These appear to play no role at two loops (they likely do appear
at higher loop order) since
the symbol of the $n$-point two-loop MHV amplitude
was computed
in~\cite{CaronHuot:2011ky} and nothing more exotic than the
examples shown in eq.~(\ref{eq:aexample}) occurs.

We emphasize that not every homogeneous polynomial of Pl\"ucker coordinates
is an $\mathcal{A}$-coordinate, nor is every cross-ratio one can write
down an $\mathcal{X}$-coordinate.   The only surefire algorithm for determining
such coordinates is via the mutation algorithm (see~\cite{Golden:2013xva}),
but we note here an empirical rule for selecting $\mathcal{X}$-coordinates
for which we know no counterexample:  a conformally invariant
ratio $x$ of $\mathcal{A}$-coordinates is
an $\mathcal{X}$-coordinate if $1+x$ also factors into a ratio of products
of $\mathcal{A}$-coordinates and if $x$ is positive-valued everywhere
inside the positive domain (this is the subset of $\Conf_n(\mathbb{P}^3)$
for which $\ket{abcd} > 0$ whenever $a<b<c<d$)\footnote{It is a logical
possibility that there could exist some $x$ which satisfies this criterion yet which is not
an $\mathcal{X}$-coordinate, though we have never encountered such an object in various
explorations through $n=9$.}.
This algorithm reveals for example that between
\begin{equation}
\frac{\ket{1235} \ket{1278} \ket{2456} \ket{5678}}{\ket{1256} \ket{2578} \ket{78 (123) \cap (456)}}
\quad \text{and}
\quad
- \frac{\ket{2(13)(56)(78)} \ket{5(12)(46)(78)}}{\ket{1256} \ket{2578} \ket{78(123) \cap (456)}}
\end{equation}
(whose difference is 1) only the first is an $\mathcal{X}$-coordinate.

\subsection{Cluster polylogarithm functions}

Now we turn to the heart of the paper: providing a definition of \emph{cluster polylogarithm functions}.
Good definitions in mathematics must lie in a Goldilocks zone: they must be sufficiently constraining
so as to select out only certain objects with interesting behavior, yet they must not be
so constraining as to preclude the existence of any examples.
In defining cluster polylogarithm functions we are guided by the physics of two-loop MHV amplitudes
in SYM theory: these functions certainly exist, yet have properties which render them very special
amongst the class of all weight-four polylogarithm functions on $\Conf_n(\mathbb{P}^3)$.

We first define a cluster $\mathcal{A}$-function of weight $k$ to
to be a conformally invariant function of transcendentality weight $k$
whose symbol can be written with only the
$\mathcal{A}$-coordinates of some cluster algebra appearing in its entries.
Functions of this type for the $\Gr(4,6)$ cluster algebra (and satisfying
various other physical constraints) were extensively classified
and studied in the papers~\cite{Dixon:2011pw,Dixon:2011nj,Dixon:2013eka}.

Our goal here is to impose additional mathematical constraints to focus
on a different (and at least for larger $n$, much smaller)
collection of functions: those which ``depend on'' only the
cluster $\mathcal{X}$-coordinates of some cluster algebra.
At weight $k<4$, where we know that the classical polylogarithm functions
generate all of $\mathcal{L}_k$, we can make this statement immediately
precise:  a cluster $\mathcal{X}$-function
of weight $k<4$ is a linear
combination of the functions $-\Li_k(-x)$ for $x$ drawn from the set
of $\mathcal{X}$-coordinates of some cluster algebra.

At weight 1 there is no distinction between
cluster $\mathcal{A}$- and $\mathcal{X}$-functions because any conformally
invariant cross-ratio can be expressed as a ratio of products
of $\mathcal{X}$-coordinates.  Hence any conformally invariant
linear combination of logarithms of $\mathcal{A}$-coordinates can be reexpressed
as a linear combination of logarithms of $\mathcal{X}$-coordinates.

At weight 2 there is still no distinction;
cluster $\mathcal{A}$-functions consist of all functions $-\Li_2(-y)$
for which both $y$ and $1+y$ factor into ratios of products of
$\mathcal{A}$-coordinates.
But then either $y$ or $-(1+y)$ (whichever is positive throughout the positive
domain)
is a cluster $\mathcal{X}$-coordinate by the criterion discussed above.
If $y$ is not the $\mathcal{X}$-coordinate then we can represent the function
$-\Li_2(-y)$ equivalently by $\Li_2(1+y)$ (modulo $\pi^2$ and products of $\log$s), establishing
that it is a cluster $\mathcal{X}$-function.

At weight 3 there is a third term in the polylogarithm identity
\begin{equation}
- \Li_3(-y) - \Li_3(1+y) - \Li_3(1+1/y) = 0 ~ (\text{modulo products of logs}),
\end{equation}
which implies that ``only half'' of weight-3 cluster $\mathcal{A}$-functions
are $\mathcal{X}$-functions.
More precisely:  if $m$ is the dimension of the space spanned by
the functions $-\Li_3(-x)$
for all cluster $\mathcal{X}$-coordinates $x$, then the space of weight-3
cluster $\mathcal{A}$-functions is $2m$ dimensional, containing in
addition all functions of the form $-\Li_3(1+x)$.

Weight 4 is the first place where things become nontrivial.
We first need a more precise definition of cluster $\mathcal{X}$-functions,
since not every weight-four polylogarithm can be expressed in terms of the
classical function $\Li_4$ only.  Motivated by the results
of~\cite{Golden:2013xva} we define a weight-four cluster $\mathcal{X}$-function
(henceforth referred to simply as a \emph{cluster polylogarithm function}
or just \emph{cluster function}) to be a cluster $\mathcal{A}$-function
whose coproduct components can be written as a linear
combination of $\{x_i\}_2 \wedge \{x_j\}_2$ or $\{x_i \}_3 \otimes x_j$ for cluster
$\mathcal{X}$-coordinates $x_i, x_j$.
Of course the classical function $-\Li_4(-x)$ is trivially such a
cluster $\mathcal{X}$-function whenever $x$ is an $\mathcal{X}$-coordinate,
so we will often use the word ``nontrivial'' to denote those
cluster $\mathcal{X}$-functions with nonzero $\Lambda^2 \B_2$ content.

We do not yet propose a definition of cluster functions for
weight greater than 4.
As mentioned above,
an appropriate definition would be as restrictive as possible
without ruling out the existence of non-trivial examples,
and should include interesting examples of functions from
SYM theory.
We believe that the identification of a suitable definition
requires first a better understanding of the structure of
MHV amplitudes
at higher loop order, of which the only example currently
in the literature is the tour de force calculation
of the three-loop MHV amplitude for $n=6$ in~\cite{Dixon:2013eka}.

In the next two sections we classify and study the properties
of the cluster functions for the simplest nontrivial cluster algebras.

\section{The $A_2$ function}
\label{sec:3}

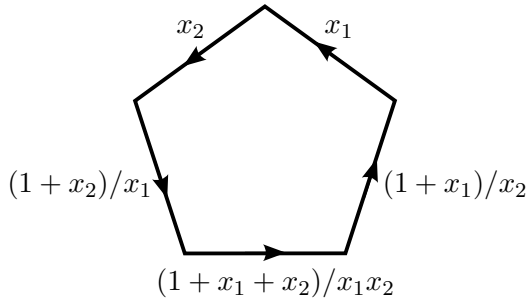
\begin{figure}
\begin{center}
\begin{picture}(100,130)
\put(0,15){\includegraphics[width=100pt]{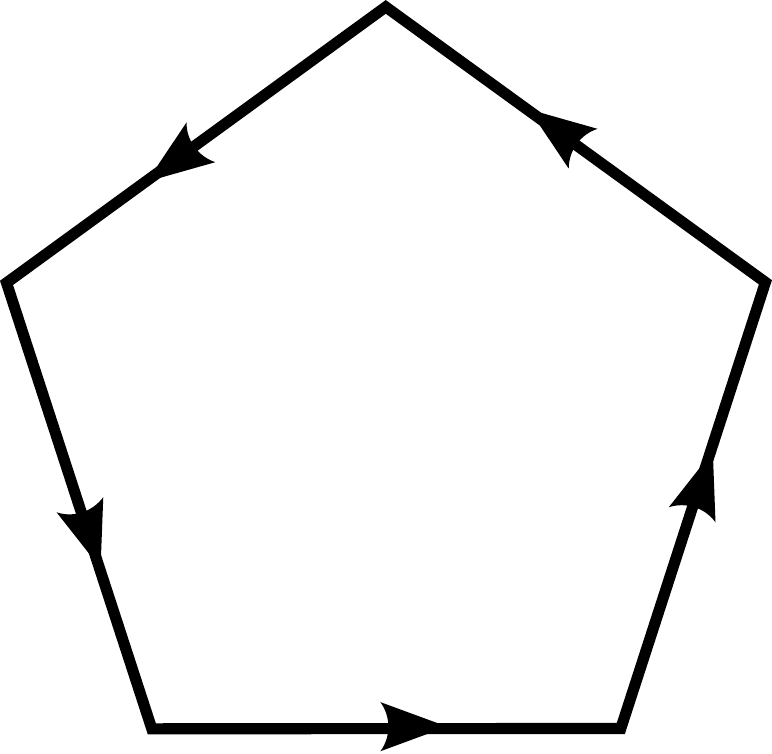}}
\put(47,0){\makebox(14,14){$(1+x_1+x_2)/x_1 x_2$}}
\put(115,37){\makebox(14,14){$(1+x_1)/x_2$}}
\put(71,96){\makebox(14,14){$x_1$}}
\put(15,96){\makebox(14,14){$x_2$}}
\put(-27,37){\makebox(14,14){$(1+x_2)/x_1$}}
\end{picture}
\end{center}
\caption{The $A_2$ cluster algebra:
to each oriented edge is associated a cluster $\mathcal{X}$-variable (reversing an arrow
requires inverting the associated variable), and to each vertex
is associated the pair of variables (called the cluster) associated to the edges emanating from that vertex.
Moving from one cluster to an adjacent one along some edge
is accomplished by mutating on the variable
associated to that edge.}
\label{fig:A2}
\end{figure}

Let us begin with the simplest nontrivial cluster algebra, the
$\Gr(3,5)$ (or $A_2$) algebra.  This algebra has five cluster
$\mathcal{X}$-coordinates which may be generated from an initial
pair $x_1, x_2$ via the relation
\begin{equation}
\label{eq:A2relation}
x_{i+1} = \frac{1 + x_i}{x_{i-1}}.
\end{equation}
Several relevant pieces of information about this algebra are
encoded graphically in the pentagon shown in fig.~\ref{fig:A2}.
To each oriented edge is associated a cluster $\mathcal{X}$-variable $x$;
in each case
$1/x$ would be associated
to the same edge with opposite orientation.
To each vertex is associated the pair of variables (the \emph{cluster})
given by the edge variables emanating away from that vertex---so, for example,
the cluster associated with the top vertex in the figure contains
the variables $(x_2, 1/x_1)$.

We seek nontrivial cluster polylogarithm functions of weight 4---that is,
solutions of eq.~(\ref{eq:integrability}) for which $b_{22}$ and $b_{31}$
can be written simply in terms of
the five available cluster $\mathcal{X}$-coordinates.
Since $A_2$ is a finite cluster algebra, this is a simple problem
in linear algebra.  The dimension of $B_1$ is 5---spanned
by the five multiplicatively independent $\mathcal{X}$-coordinates,
the dimension of $\B_2$ is 4---spanned by the five $\{x_i\}_2$ subject
to the Abel identity
\begin{equation}
\sum_{i=1}^5 \{ x_i \}_2 = 0,
\end{equation}
and the dimension of $\B_3$ is again 5---spanned by the five
$\{x_i\}_3$, which are independent.

It is simple to check that in the 10-dimensional
space $\Lambda^2 \B_2$,
there is a unique element $b_{22}$
for which there exists a $b_{31}$ in the 25-dimensional $\B_3 \otimes \mathbb{C}^*$
satisfying eq.~(\ref{eq:integrability}).
We call this solution the \emph{$A_2$ function} (or the
pentagon function).  The $\B_3 \otimes \mathbb{C}^*$ component of
the $A_2$ function is not uniquely fixed by eq.~(\ref{eq:integrability})
since one always has the freedom to add any linear combination of the five
$-\Li_4(-x_i)$.  We fix this freedom by choosing to define
the $A_2$ function to have the coproduct components
\begin{equation}
\begin{aligned}
\delta f_{A_2}(x_1,x_2)\rvert_{\Lambda^2 \B_2} &=\sum_{i,j=1}^5 j \{x_i\}_2 \wedge \{x_{i+j}\}_2, \\
\delta f_{A_2}(x_1,x_2)\rvert_{\B_3 \otimes \mathbb{C}^*}
&=5\sum_{i=1}^5\left( \{x_{i+1}\}_3 \otimes x_{i}-\{x_i\}_3 \otimes x_{i+1}\right).
\label{eq:pentagonmotivic}
\end{aligned}
\end{equation}
This is the unique choice which is skew-dihedral invariant---that means it is (1) cyclically
invariant under $x_i \to x_{i+1}$ and (2) changes sign under
$x_i \to x_{6-i}$.
A very important facet of this definition is the antisymmetry of $\delta f_{A_2}(x_1,x_2)\rvert_{\B_3 \otimes \mathbb{C}^*}$ under $\{x\}_3\otimes y \to \{y\}_3\otimes x$.
In some sense we can therefore consider $f_{A_2}$ to be a ``purely
non-classical'' cluster function (although this notion is not
precisely defined),
since any linear combination of the classical functions $-\Li_4(-x_i)$ functions has a naturally symmetric
$\B_3 \otimes \mathbb{C}^*$ component.
This antisymmetry property of the $A_2$ function
makes them useful building blocks for
expressing scattering amplitudes,
as discussed below in sec.~\ref{sec:5}.

It is also interesting to note that the $\B_3 \otimes \mathbb{C}^*$ content of $f_{A_2}$ can be expressed
in an evidently ``local'' manner---by this we mean that the two $\mathcal{X}$-coordinates
in each term $\{x_i\}_3 \otimes x_j$ always have $j=i\pm 1$ and therefore appear together inside some cluster and moreover have Poisson bracket $\{ x_i, x_{i\pm 1}\}= \pm 1$.
In contrast, the $\Lambda^2 \B_2$ component is \emph{non}-local:
the two variables appearing in each term $\{x_i\}_2 \wedge \{x_j\}_2$ do not in general
appear together in a common cluster and do not have
any particularly simple Poisson bracket with each other.

Let us pause to clarify one point of notation which will allow us to avoid confusion later.
All five $\mathcal{X}$-coordinates appear on the right-hand sides of~(\ref{eq:pentagonmotivic}),
but we appropriately write $f_{A_2}(x_1,x_2)$ as a function of only two variables since
the others may be expressed in terms of these via the relation~(\ref{eq:A2relation}).
Below we will frequently need to discuss $A_2$ subalgebras of larger cluster algebras.
Any such subalgebra is generated by a pair of $\mathcal{X}$-coordinates which appear together inside some cluster and which have Poisson
bracket $\{x,y\} = 1$.  When this happens the corresponding $A_2$ function
is simply $f_{A_2}(x,y)$.
To summarize using the quiver notation reviewed in~\cite{Golden:2013xva}:
$f_{A_2}(x,y)$ is a function of any two $\mathcal{X}$-coordinates
appearing inside a quiver as $x \to y$.

We emphasize that the equations~(\ref{eq:pentagonmotivic}) completely and
unambiguously define the $A_2$
function as an element of $\mathcal{L}_4$---i.e., modulo products of functions of lower weight.
Nevertheless, the reader with an appetite for seeing an \emph{actual function} with these
coproduct components may turn to the appendix for satisfaction, and we can write here
a relatively simple expression for the symbol of a representative of $f_{A_2}$:
\begin{equation}
\label{eq:A2symbol}
\text{symbol}(f_{A_2}(x_1,x_2)) \sim \frac{5}{4} \sum_{\text{skew-dihedral}}
x_1 \otimes x_2 \otimes x_1 \otimes \frac{x_2}{x_5} + x_1 \otimes x_2 \otimes x_2 \otimes \frac{x_1}{x_3}.
\end{equation}
We write $\sim$ instead of $=$ because as long as we consider $f_{A_2}$ only as an element
of $\mathcal{L}_4$ its symbol is not even well-defined---eq.~(\ref{eq:A2symbol}) represents one particular
way of fixing the ambiguity associated to products of lower-weight
functions (it is the choice which makes the symbol an eigenvector of $\rho$), but we are not
yet ready to commit to any choice.

Although we believe this function to be new (and hopefully interesting) to the mathematics
community, it may seem that this example is too trivial to be relevant to SYM theory, where the
relevant algebras are $\Gr(4,n)$.  For sure, $\Gr(4,n)$ contains many $A_2$ subalgebras, and
we may evaluate $f_{A_2}$ on each of these, but are there any other solutions
of~(\ref{eq:integrability}) for these algebras?  Surprisingly, we have checked in
addition to $A_2$ the finite
algebras $A_3$, $A_4$ and $D_4$,
and in each case we have found that there are no other
solutions---for these cluster algebras, \emph{all non-trivial weight-four cluster functions are
linear combinations of $A_2$ functions}\footnote{This statement has also been verified for the
$E_6$ ($=\Gr(4,7)$) cluster algebra by D.~Parker and A.~Scherlis.}!

It remains an interesting mathematical problem to determine, for general cluster algebras (even for infinite ones),
the set of non-trivial cluster polylogarithm functions; that is, the subspace of $\Lambda^2 \B_2$ on
which~(\ref{eq:integrability}) can be
solved in terms of
an element $b_{31}$ expressible purely in terms of cluster
$\mathcal{X}$-coordinates.  However, even if more
exotic solutions exist in general, for the limited purpose of studying two-loop $n$-point MHV amplitudes
it seems clear that the $A_2$ functions are completely sufficient, in part because these amplitudes
only live in a finite (and small) piece of the relevant cluster algebras, as discussed in sec.~\ref{sec:5}.

\section{The $A_3$ function}
\label{sec:4}

We now turn our attention to cluster polylogarithms for the $A_3$ cluster algebra, beginning with the seed quiver $x_1 \to x_2 \to x_3$\footnote{Note that this is really shorthand for ``a triplet of $\mathcal{X}$-coordinates $\{x_1, x_2, x_3\}$ that are all in the same cluster (this distinguishes between $x_i$ and $1/x_i$) and have the Poisson structure $\{x_1,x_2\}=\{x_2,x_3\}=1,~\{x_1,x_3\}=0$."}. This quiver generates the following 15 cluster $\mathcal{X}$-coordinates:
\begin{alignat}{3}\label{eq:A3coords}
	x_{1,1} &= x_1  &\quad x_{1,2} &= 1/x_3 &\quad v_1 &=\frac{\left(x_2+1\right) \left(x_1 x_2 x_3+x_2 x_3+x_3+1\right)}{x_1 x_2}\nonumber \\
	x_{2,1} &= \left(x_1 x_2+x_2+1\right) x_3 & x_{2,2}
    	&=\frac{x_1 x_2+x_2+1}{x_1} & v_2 &=\frac{x_3+1}{x_2 x_3}\nonumber\\
    x_{3,1} &= \frac{x_2 x_3+x_3+1}{x_2} & x_{3,2} &= \frac{x_2 x_3+x_3+1}{x_1 x_2 x_3} & v_3 &=\left(x_1+1\right) x_2\\
    e_1 &= \frac{x_1 x_2 x_3+x_2 x_3+x_3+1}{\left(x_1+1\right) x_2} & e_2 &=\frac{1}{\left(x_2+1\right) x_3} & e_3 &= \frac{\left(x_1+1\right) x_2 x_3}{x_3+1}\nonumber\\
    e_4 &= \frac{x_2+1}{x_1 x_2} & e_5 &= \frac{x_1 \left(x_3+1\right)}{x_1 x_2 x_3+x_2 x_3+x_3+1} & e_6 &= x_2.\nonumber
\end{alignat}

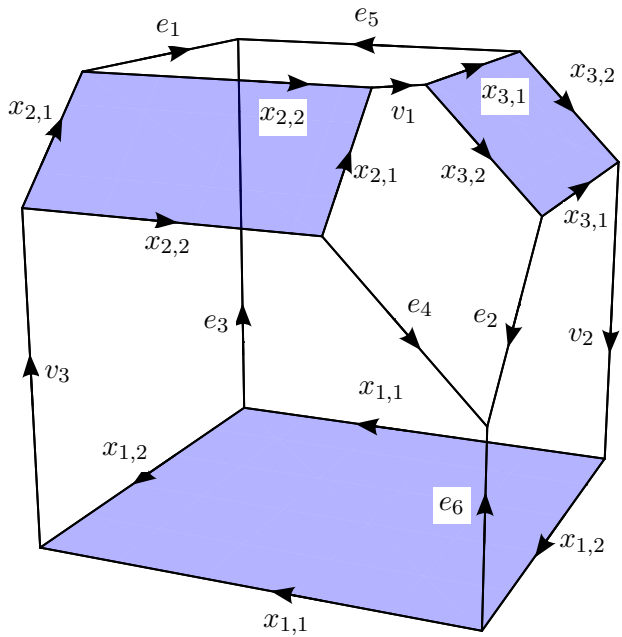
\begin{figure}
\begin{center}
\begin{picture}(240,260)
\put(0,0){\includegraphics[width=240pt]{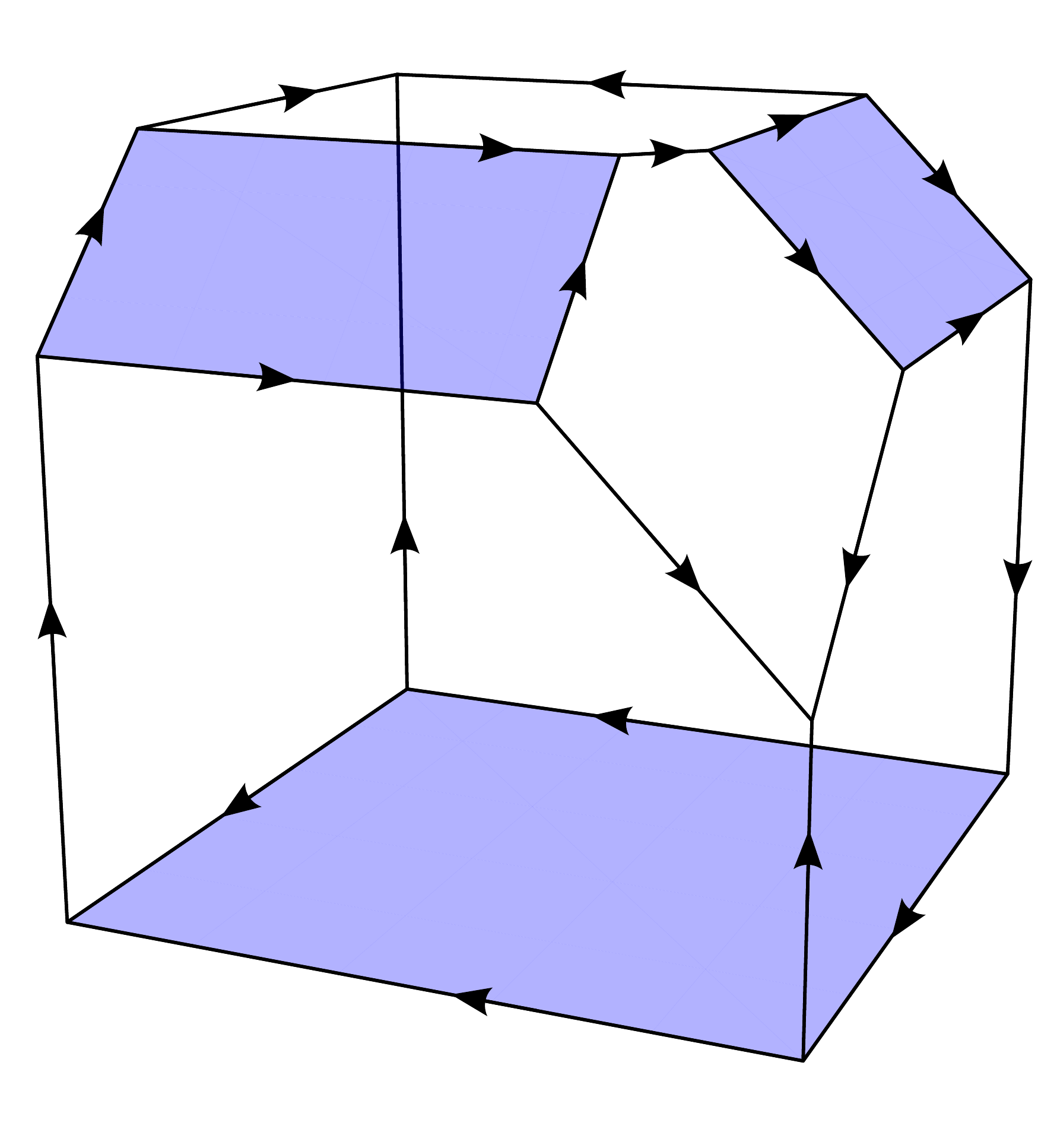}}
\put(100,15){\colorbox{white}{\makebox(17,14){$x_{1,1}$}}}
\put(212,45){\colorbox{white}{\makebox(17,14){$x_{1,2}$}}}
\put(39,80){\colorbox{white}{\makebox(17,14){$x_{1,2}$}}}
\put(136,103){\colorbox{white}{\makebox(17,14){$x_{1,1}$}}}
\put(164,60){\colorbox{white}{\makebox(14,14){$e_6$}}}
\put(75,129){\colorbox{white}{\makebox(14,14){$e_3$}}}
\put(15,110){\colorbox{white}{\makebox(14,14){$v_3$}}}
\put(213,124){\colorbox{white}{\makebox(14,14){$v_2$}}}
\put(55,158){\colorbox{white}{\makebox(17,14){$x_{2,2}$}}}
\put(5,209){\colorbox{white}{\makebox(14,14){$x_{2,1}$}}}
\put(134,185){\colorbox{white}{\makebox(17,14){$x_{2,1}$}}}
\put(167,185){\colorbox{white}{\makebox(17,14){$x_{3,2}$}}}
\put(216,223){\colorbox{white}{\makebox(17,14){$x_{3,2}$}}}
\put(98,207){\colorbox{white}{\makebox(18,14){$x_{2,2}$}}}
\put(213,168){\colorbox{white}{\makebox(18,14){$x_{3,1}$}}}
\put(152,135){\colorbox{white}{\makebox(14,14){$e_4$}}}
\put(177,131){\colorbox{white}{\makebox(14,14){$e_2$}}}
\put(146,208){\colorbox{white}{\makebox(14,14){$v_1$}}}
\put(57,241){\colorbox{white}{\makebox(14,14){$e_1$}}}
\put(182,215){\colorbox{white}{\makebox(18,14){$x_{3,1}$}}}
\put(132,245){\colorbox{white}{\makebox(14,14){$e_5$}}}
\end{picture}
\end{center}
\caption{The Stasheff polytope for the $A_3$ cluster algebra.  The caption of fig.~(\ref{fig:A2})
applies, except that here a cluster of three
$\mathcal{X}$-coordinates is associated to each vertex.  The three quadrilateral faces
are shaded blue to distinguish them visually from the six pentagonal faces.
The interior of this polytope can be identified with the blow-up of the positive domain
in $\Conf_6(\mathbb{P}^3)$, see for example~\cite{Brown}.}
\label{fig:A3}
\end{figure}

The structure of the algebra is summarized in the Stasheff polytope shown in fig.~(\ref{fig:A3}).
The polytope has 9 faces (comprising six pentagons and three quadrilaterals),
14 vertices, and 21 edges, each of which is labeled by an $\mathcal{X}$-coordinate.

We now review a few facts about the natural Poisson structure~\cite{FG03b} on $\Conf_n(\mathbb{P}^3)$
following~\cite{Golden:2013xva}.
A pair of cluster $\mathcal{X}$-coordinates has a simple Poisson bracket (``simple'' means
$\pm 1$ or 0) only if they appear together inside some cluster.
The coordinates in eq.~(\ref{eq:A3coords}) have the following Poisson structure:
\begin{equation}
        \{x_{i,1},x_{i,2}\}=0,\quad\{e_i,e_{i+4}\}=1,\quad\{v_i,x_{i\pm1,a}\}=\mp1,\quad\{e_i,x_{i+1,a}\}=-1,
\end{equation}
where $v$ and $x$ have indices mod 3 and $e$ has indices mod 6. This means that there are 3 pairs of $\mathcal{X}$-coordinates that Poisson commute and 30 pairs with Poisson bracket $\pm1$.

Quadrilateral faces of a Stasheff polytope are in correspondence
with pairs of cluster $\mathcal{X}$-coordinates which Poisson commute, thereby generating
$A_1 \times A_1$ subalgebras.
Pentagonal faces of a Stasheff polytope correspond to $A_2$ subalgebras, generated by pairs of
cluster $\mathcal{X}$-coordinates which have Poisson bracket $\pm 1$.
For the $A_3$ algebra there are 30 such pairs---5 (one at each vertex) each for the six
pentagonal faces evident in fig.~(\ref{fig:A3}).
The sign of the Poisson bracket is unfortunately not manifest in the figure, so we record
here explicitly
the five $\mathcal{X}$-coordinates $(x_1,\ldots,x_5)$ (following the notation of
sec.~\ref{sec:3}) for each of the six $A_2$ subalgebras:
\begin{alignat}{4}
\vcenter{\hbox{\includegraphics[width=40pt]{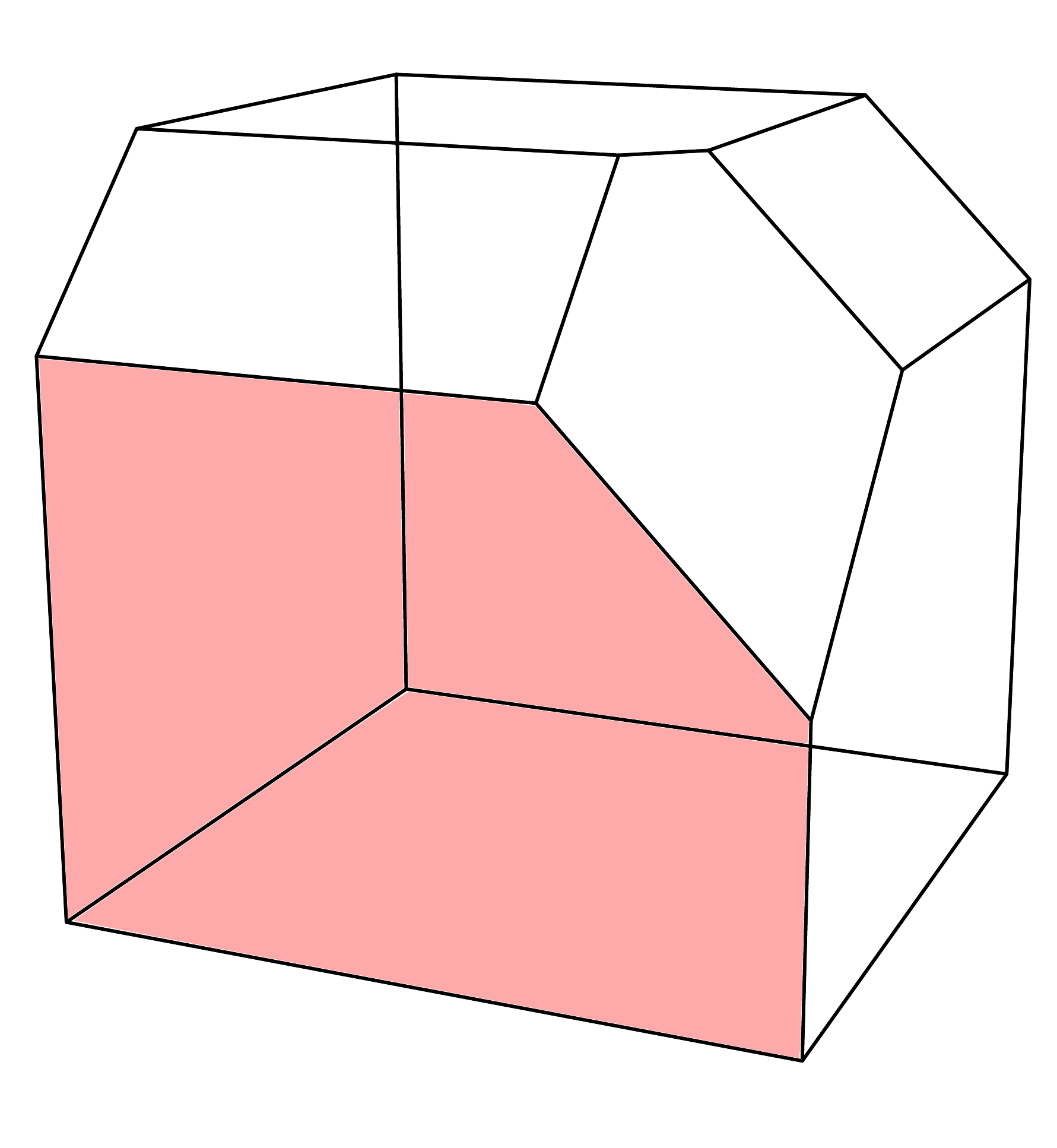}}}
& \quad (e_4,1/e_6,x_{1,1},v_3,x_{2,2}),&
\vcenter{\hbox{\includegraphics[width=40pt]{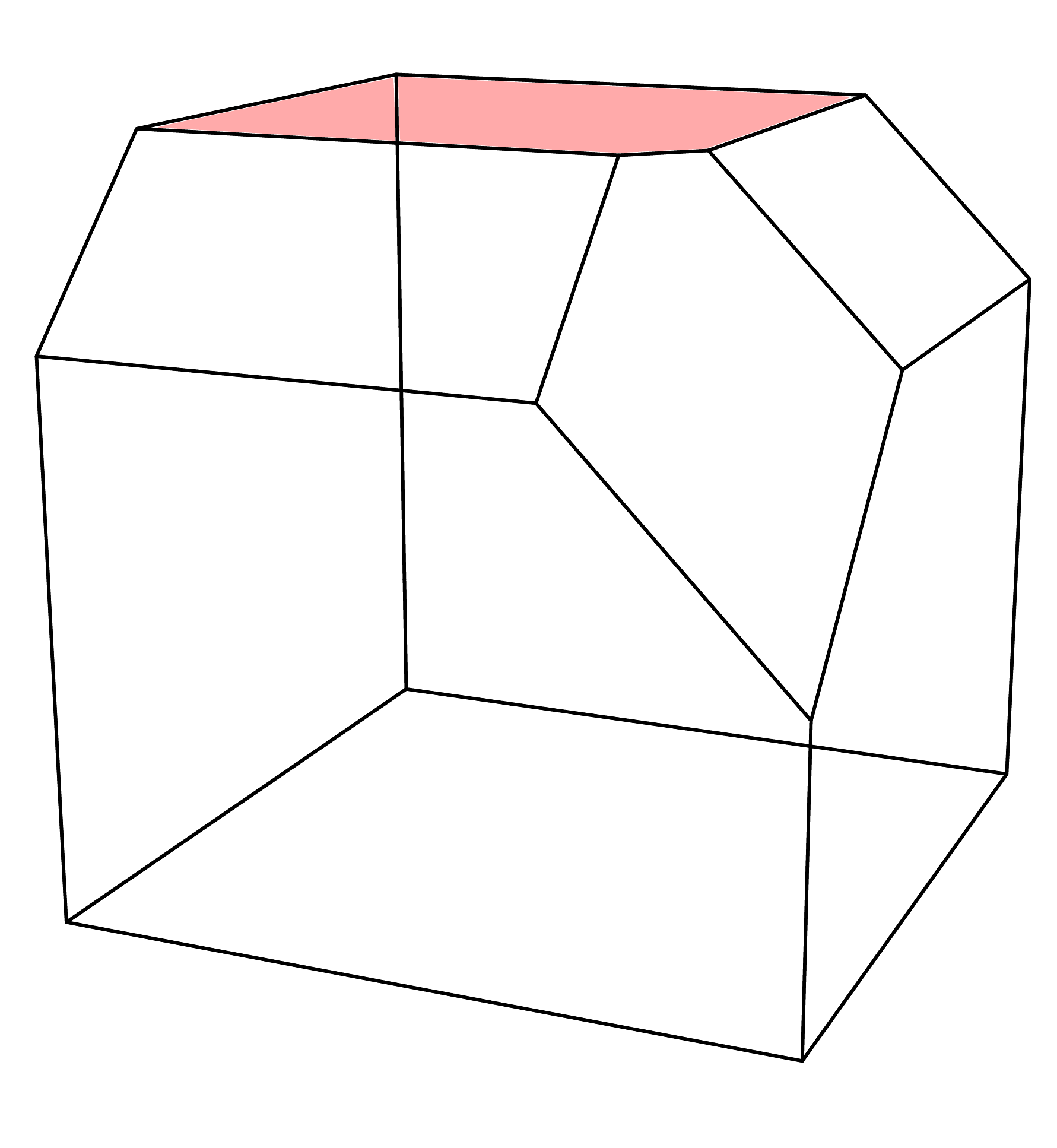}}}
& \quad (e_5,1/e_1,x_{2,2},v_1,x_{3,1}),&\nonumber\\
\vcenter{\hbox{\includegraphics[width=40pt]{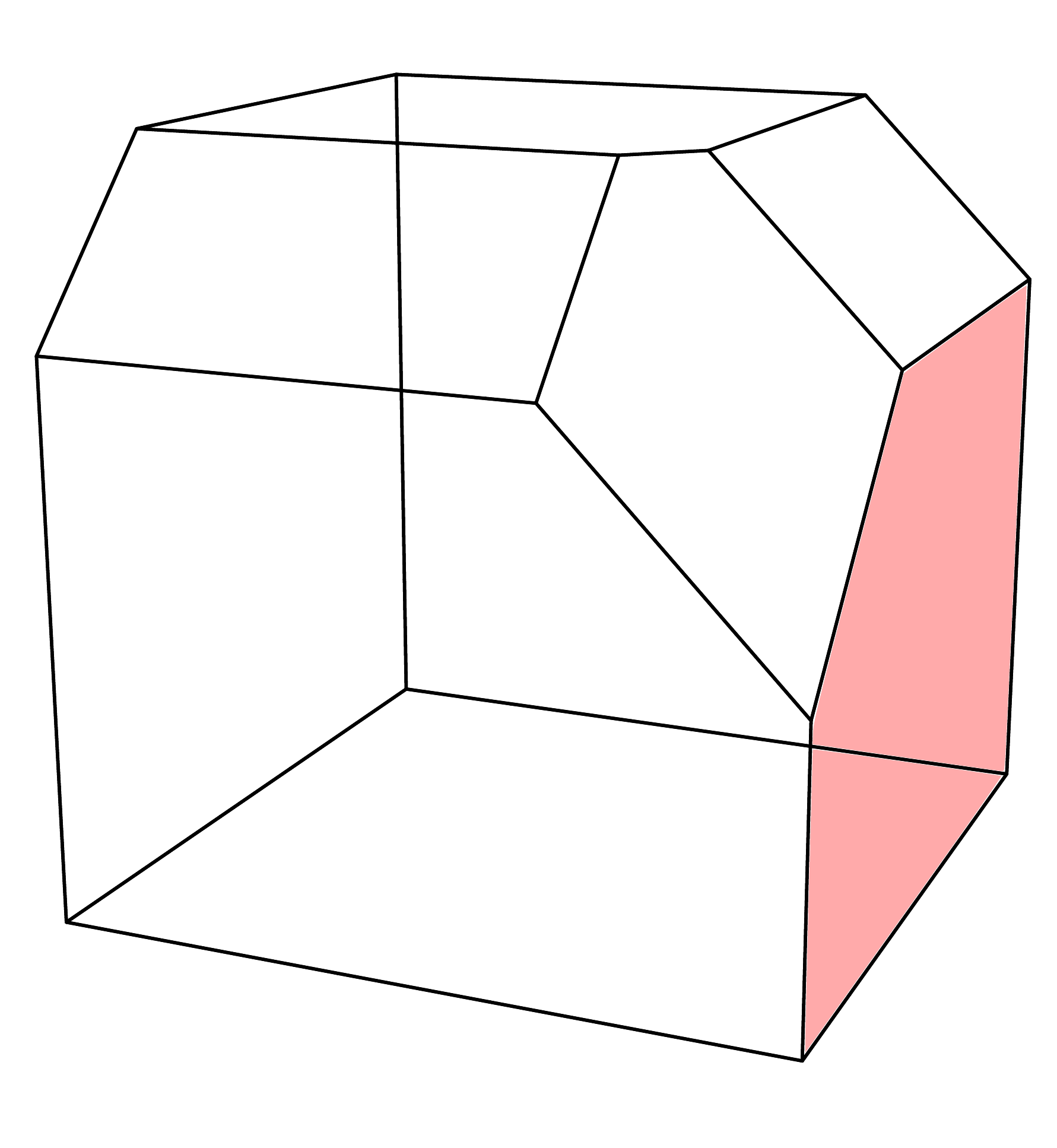}}}
& \quad (e_6,1/e_2,x_{3,1},v_2,x_{1,2}),\qquad &
\vcenter{\hbox{\includegraphics[width=40pt]{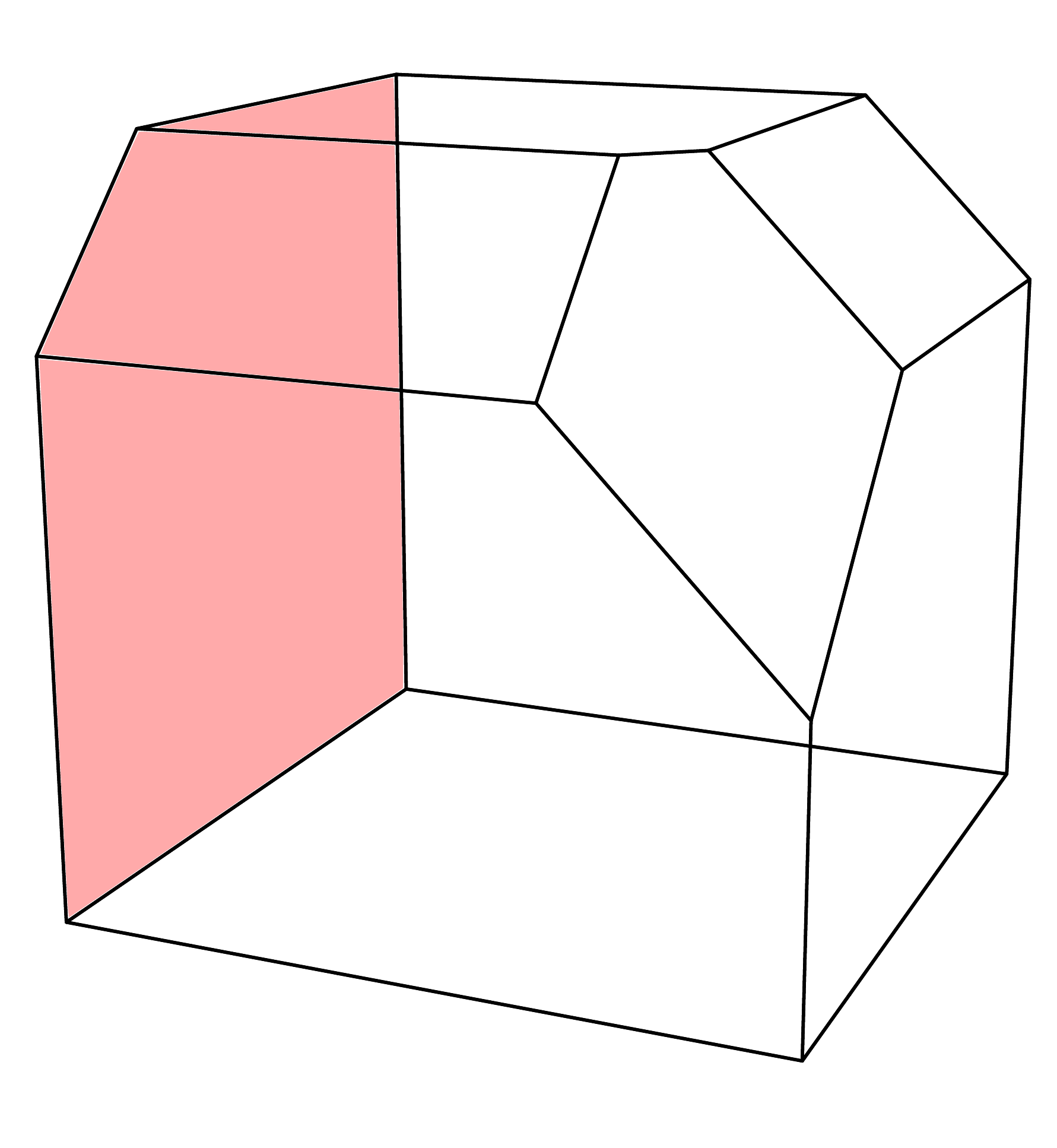}}}
& \quad (e_1,1/e_3,x_{1,2},v_3,x_{2,1}),& \label{eq:pentagons}\\
\vcenter{\hbox{\includegraphics[width=40pt]{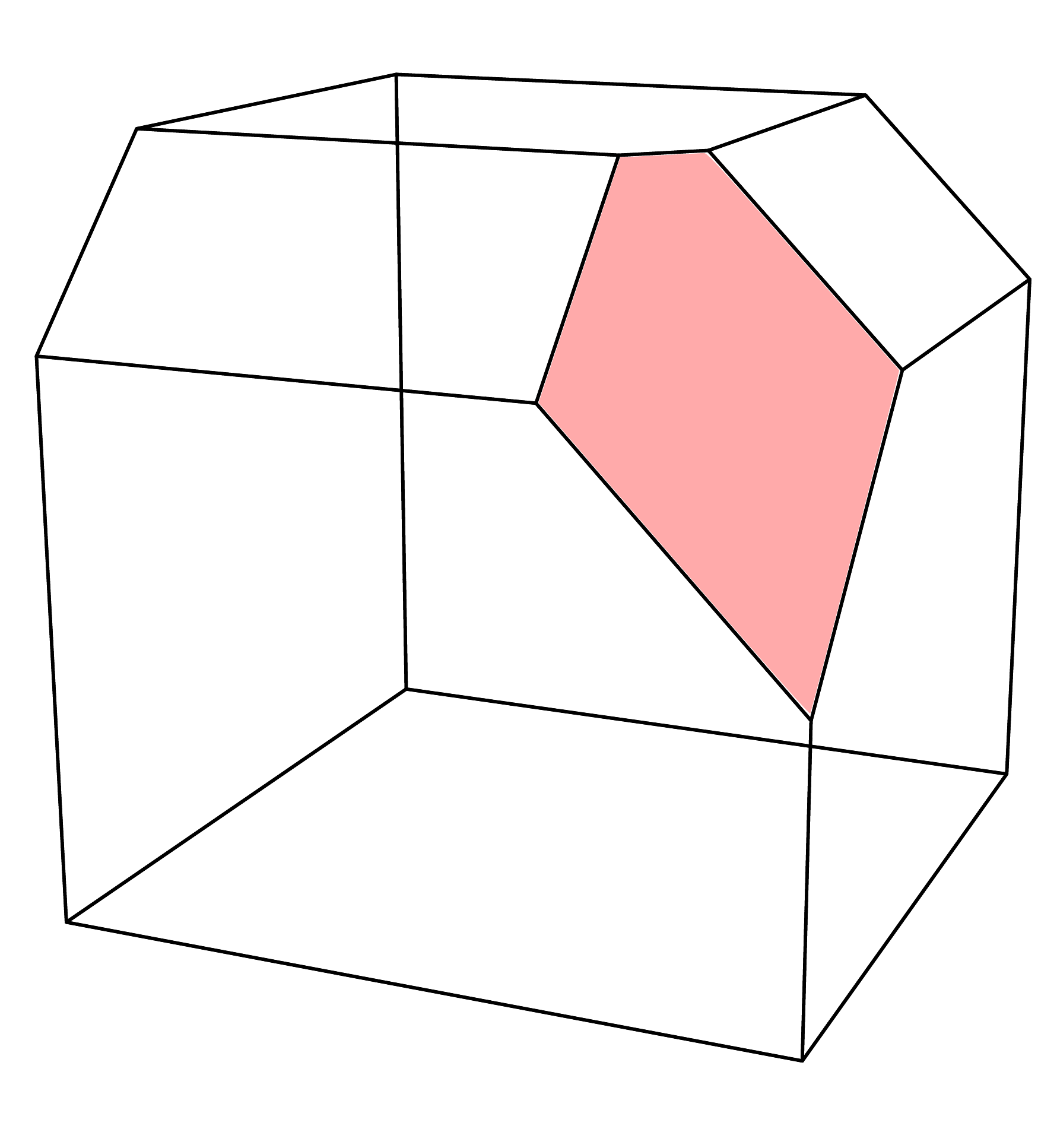}}}
& \quad (e_2,1/e_4,x_{2,1},v_1,x_{3,2}),&
\vcenter{\hbox{\includegraphics[width=40pt]{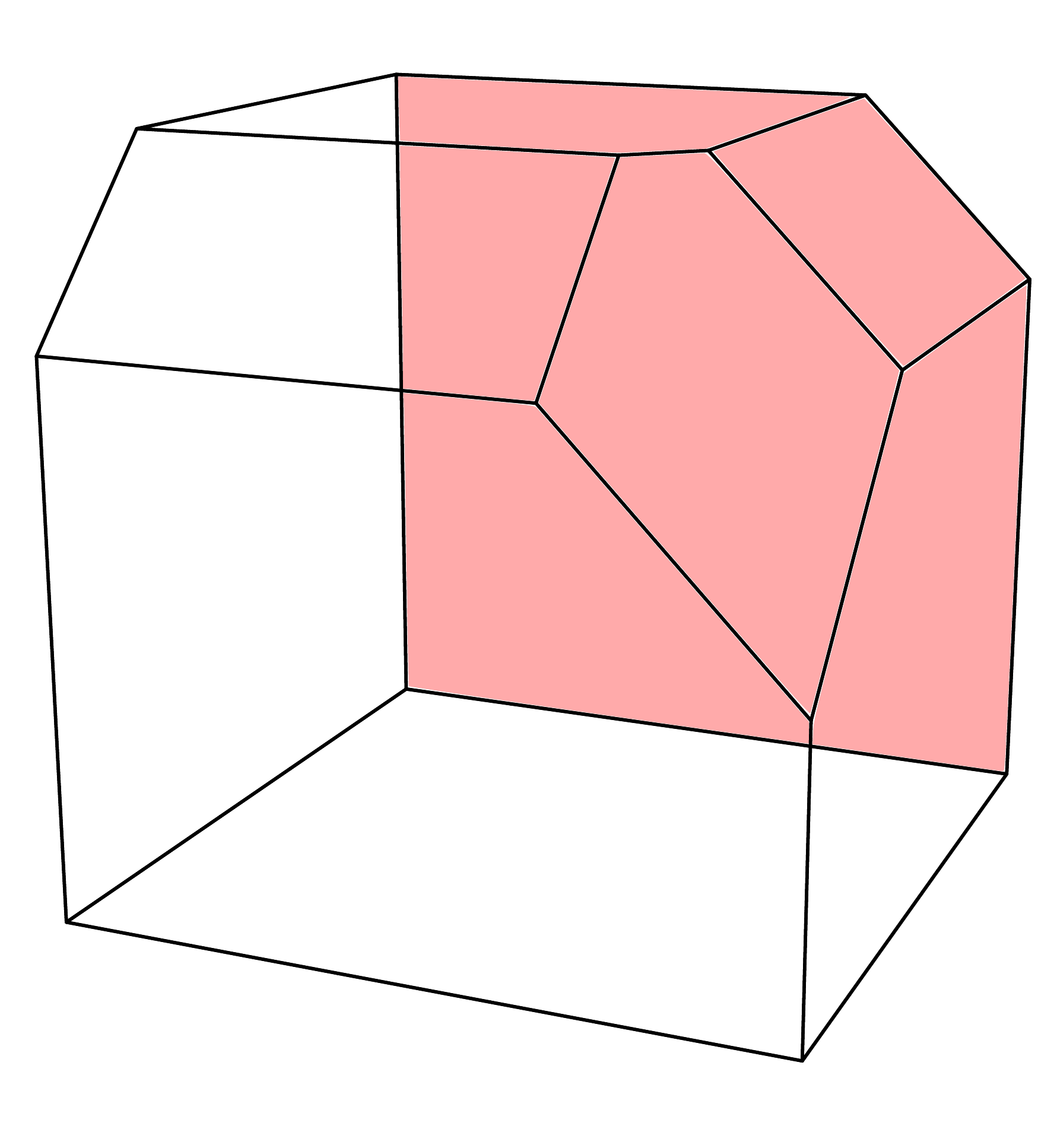}}}
& \quad (e_3,1/e_5,x_{3,2},v_2,x_{1,1}).&\nonumber
\end{alignat}
Each cyclically adjacent pair of variables appearing here,
for example $\{e_6,1/e_4\}$ or $\{x_{2,1},e_1\}$, has Poisson bracket $+1$.
The three entries in the left column can be read off from fig.~(\ref{fig:A3}) by going around
the pentagons clockwise (as seen from outside the Stasheff polytope), while the three entries in
the right column must be read off
counterclockwise.

Finally we come to the question of cluster functions for the $A_3$ algebra.  As revealed
already at the end of the previous section, it is a simple problem in linear algebra to
verify that the equation~(\ref{eq:integrability}) admits solutions only when $b_{22}$
lies in the 6-dimensional
subspace of $\Lambda^2 \B_2$ spanned by the six $A_2$ functions associated to~(\ref{eq:pentagons}).
We may represent these six functions
as $f_{A_2}(e_i,1/e_{i+2})$ for $i=1,\ldots,6$ thanks to the cyclic invariance of the $A_2$
function.

It is now time, in our quest to cook up a fine selection of special functions for the two-loop MHV amplitudes,
to toss in one more very special ingredient.  Beyond the fact that they are cluster polylogarithm
functions, an even more amazing property of these amplitudes is that they have $\Lambda^2 \B_2$ content
which can be expressed entirely in terms of pairs of cluster $\mathcal{X}$-coordinates
$\{x_i\}_2 \wedge \{x_j\}_2$ which Poisson commute:  $\{x_i,x_j\} = 0$! This was shown to be true for $n=7$ in~\cite{Golden:2013xva}, and is in fact known to be true for all $n$~\cite{futureWork,SimonsTalk}.

For the $A_3$ algebra it is simple
to check that there is a unique linear combination of the six $A_2$ functions with this property,
which we naturally call the $A_3$ function:
\begin{equation}
\label{eq:A3definition}
f_{A_3} = \frac{1}{10} \sum_{i=1}^6 (-1)^i f_{A_2}(e_i,1/e_{i+2}).
\end{equation}
The coproduct of the $A_3$ function has the spectacularly simple, ``local'' $\Lambda^2 \B_2$ content
\begin{equation}\label{eq:A3motivic}
        \delta f_{A_3}|_{\Lambda^2 B_2} = \sum_{i=1}^3 \{x_{i,1}\}_2 \wedge \{x_{i,2} \}_2.
\end{equation}
We do not write the $\B_3 \otimes \mathbb{C}^*$ component since it does not simplify beyond
the alternating sum of six copies of the corresponding component from the $A_2$
function.

We observed beneath eq.~(\ref{eq:pentagonmotivic}) that the $\B_3 \otimes \mathbb{C}^*$
content of the $A_2$ function is ``local'' (involving only pairs of variables which appear
in a common cluster), and the $A_3$ function obviously inherits this property.
However the $A_2$ function has a non-local $\Lambda^2 \B_2$ component, so it is rather amazing
that the particular linear combination of $A_2$'s appearing inside $A_3$ give rise to the completely
local eq.~(\ref{eq:A3motivic}).
Moreover, the two coproduct components see distinct aspects of the geometry of the Stasheff
polytope---the $\Lambda^2 \B_2$ component involves the three quadrilateral
faces (i.e., the $A_1 \times A_1$ subalgebras)
while the $\B_3 \otimes \mathbb{C}^*$ component involves the six pentagonal faces
(the $A_2$ subalgebras).
It is tempting to anticipate the possibility that this notion of locality within the Stasheff
polytope might underlie the structure of SYM theory scattering amplitudes in a very deep way.
If this proves to be so,
we cannot help but wonder (following somewhat the motivation espoused
by~\cite{ArkaniHamed:2012nw}) whether there exists an alternative
formulation of SYM theory scattering amplitudes which makes this ``locality in the Stasheff
polytope'' manifest.

A conjecture central to our approach is that the set of $f_{A_3}$
for all possible $A_3$ subalgebras of $\Gr(4,n)$ spans the space
of all weight-four cluster polylogarithm functions whose coproduct components are completely
``local'' (involving only quadrilaterals in $\Lambda^2 \B_2$ and only
pentagons in $\B_3 \otimes \mathbb{C}^*$).

We now display a simple realization of the $A_3$ function in a familiar setting: the $\Gr(4, 6)$ algebra, relevant to 6-particle scattering, which is in fact isomorphic to $A_3$. In order to align with the notation in~\cite{Golden:2013xva}, we consider $(x_1,x_2,x_3) = (x_1^-, e_6, 1/x_1^+)$ and relate $x_{i,1} = x_i^-$ and $x_{i,2} = x_i^+$. The 15 $\mathcal{X}$-coordinates can then be written as
\begin{alignat}{3}
v_1 &= \frac{\ket{1246}\ket{1345}}{\ket{1234}\ket{1456}},
&\qquad v_2 &= \frac{\ket{1235}\ket{2456}}{\ket{1256}\ket{2345}},
&\qquad v_3 &= \frac{\ket{1356}\ket{2346}}{\ket{1236}\ket{3456}},\nonumber \\
x^+_1 &= \frac{\ket{1456}\ket{2356}}{\ket{1256}\ket{3456}},
&\qquad x^+_2 &= \frac{\ket{1346}\ket{2345}}{\ket{1234}\ket{3456}},
&\qquad x^+_3 &= \frac{\ket{1236}\ket{1245}}{\ket{1234}\ket{1256}},\nonumber \\
x^-_1 &= \frac{\ket{1234}\ket{2356}}{\ket{1236}\ket{2345}},
\label{eq:ratios}
&\qquad x^-_2 &= \frac{\ket{1256}\ket{1346}}{\ket{1236}\ket{1456}},
&\qquad x^-_3 &= \frac{\ket{1245}\ket{3456}}{\ket{1456}\ket{2345}}, \\
e_1 &= \frac{\ket{1246}\ket{3456}}{\ket{1456}\ket{2346}},
&\qquad e_2 &= \frac{\ket{1235}\ket{1456}}{\ket{1256}\ket{1345}},
&\qquad e_3 &= \frac{\ket{1256}\ket{2346}}{\ket{1236}\ket{2456}}, \nonumber\\
e_4 &= \frac{\ket{1236}\ket{1345}}{\ket{1234}\ket{1356}},
&\qquad e_5 &= \frac{\ket{1234}\ket{2456}}{\ket{1246}\ket{2345}},
&\qquad e_6 &= \frac{\ket{1356}\ket{2345}}{\ket{1235}\ket{3456}}.\nonumber
\end{alignat}
Notably absent from this list are the three cross-ratios $u_1,u_2,u_3$ often used in the physics
literature; these are related to the $v_i$'s by $u_i = 1/(1 + v_i)$. Evaluating eq.~(\ref{eq:A3definition}) on the variables in (\ref{eq:ratios}) generates what we will call ``the $\Gr(4,6)$ function".

It is interesting to note that the transformation of the $\Gr(4,6)$ function
with respect to the dihedral group acting on the 6 particles is opposite to that
of the 5-particle dihedral group acting on the $A_2$ function.
Specifically, the $\Gr(4,6)$ function is invariant under flipping particle $i$ to particle $7-i$,
but it is antisymmetric under a cyclic rotation $i \to i + 1$.
This antisymmetry is manifest for example in eq.~(\ref{eq:A3motivic})
upon noting that the $x^\pm_i$ transform under a cyclic rotation according to
\begin{equation}
x^\pm_i \to x^\mp_{i+1}.
\end{equation}

The $\Gr(4,6)$ algebra has an additional involution of order 2, called
parity in~\cite{Golden:2013xva} (it corresponds to complex conjugation
in Minkowski space kinematics), under which the $\mathcal{X}$-coordinates
transform according to
\begin{equation}
\label{eq:parity}
v_i \mapsto v_i, \qquad
x_i^\pm \mapsto x^\mp_i,
\qquad
e_i \mapsto e_{i+3}.
\end{equation}
The $\Gr(4,6)$ function is antisymmetric under this parity operation.

The fact that
MHV amplitudes are required to be fully invariant under
both parity and cyclic symmetry, yet the unique
non-classical weight four function with the right cluster properties is
antisymmetric under these symmetries, ``explains why''
the two-loop 6-particle MHV amplitude~\cite{Goncharov:2010jf}
must be expressible in terms of classical polylogarithms\footnote{An explanation
with the same flavor, but based on more physical constraints
(rather than our more mathematical constraints)
was given in~\cite{Dixon:2011nj}.}.

\section{Cluster polylogarithms for $\Gr(4,7)$ and the amplitude
$R_7^{(2)}$}
\label{sec:5}

We now demonstrate the utility of the $A_3$ function for two-loop MHV scattering
amplitudes by providing, as an illustrative example,
an explicit representation of the two-loop 7-particle MHV amplitude
(modulo products of functions of lower weight, as always).  We have carried out this
exercise for $n>7$ (where the cluster algebras $\Gr(4,n)$ are of infinite type) with
no difficulty, but we relegate a detailed analysis of
these more complicated results to a future
publication~\cite{futureWork}.

First let us take a look at the $A_2$ subalgebras.
The $\Gr(4,7) = E_6$ cluster algebra has 1071 $A_2$ subalgebras (i.e., 1071 pentagonal faces
on its generalized Stasheff polytope) on which the $A_2$ function can be evaluated,
but only 504 of these give distinct results.
We can tabulate here the 504 ``distinct $A_2$ subalgebras''
by providing their quivers, in terms of cluster $\mathcal{X}$-coordinates for $\Gr(4,7)$.  First we have
\begin{equation}
	\begin{array}{l}
 \frac{\langle 1245\rangle  \langle 1567\rangle }{\langle
   1257\rangle  \langle 1456\rangle }\to \frac{\langle
   1247\rangle  \langle 1256\rangle  \langle 1345\rangle
   }{\langle 1234\rangle  \langle 1257\rangle  \langle
   1456\rangle }, \qquad \frac{\langle 1237\rangle  \langle 1245\rangle  \langle
   4567\rangle }{\langle 2457\rangle  \langle
   1(23)(45)(67)\rangle }\to \frac{\langle 1267\rangle
   \langle 1457\rangle  \langle 2345\rangle }{\langle
   2457\rangle  \langle 1(23)(45)(67)\rangle }, \\
\end{array}
\end{equation}
and their cyclic images ($2\times7 =14$ total quivers). It suffices to take just the cyclic images because both parity and $i \to 8-i$ map this set back to itself. Next we have
\begin{equation}
	\begin{array}{l}
 \frac{\langle 1236\rangle  \langle 1245\rangle }{\langle
   1234\rangle  \langle 1256\rangle }\to \frac{\langle
   1237\rangle  \langle 1246\rangle }{\langle 1234\rangle
    \langle 1267\rangle },
     \frac{\langle 1237\rangle  \langle 1246\rangle }{\langle
   1234\rangle  \langle 1267\rangle }\to \frac{\langle
   1247\rangle  \langle 1456\rangle  \langle 2346\rangle
   }{\langle 1234\rangle  \langle 1467\rangle  \langle
   2456\rangle }, \\
 \frac{\langle 1237\rangle  \langle 1246\rangle }{\langle
   1234\rangle  \langle 1267\rangle }\to -\frac{\langle
   1247\rangle  \langle 3456\rangle }{\langle
   4(12)(35)(67)\rangle },
 \frac{\langle 1236\rangle  \langle 1345\rangle }{\langle
   1234\rangle  \langle 1356\rangle }\to \frac{\langle
   1237\rangle  \langle 1346\rangle }{\langle 1234\rangle
    \langle 1367\rangle }, \\
 \frac{\langle 1236\rangle  \langle 1567\rangle }{\langle
   1267\rangle  \langle 1356\rangle }\to \frac{\langle
   1237\rangle  \langle 1256\rangle  \langle 1346\rangle
   }{\langle 1234\rangle  \langle 1267\rangle  \langle
   1356\rangle },
 \frac{\langle 1234\rangle  \langle 1357\rangle }{\langle
   1237\rangle  \langle 1345\rangle }\to \frac{\langle
   1235\rangle  \langle 1367\rangle  \langle 3457\rangle
   }{\langle 1237\rangle  \langle 1345\rangle  \langle
   3567\rangle }, \\
 \frac{\langle 1246\rangle  \langle 1345\rangle }{\langle
   1234\rangle  \langle 1456\rangle }\to \frac{\langle
   1247\rangle  \langle 1346\rangle }{\langle 1234\rangle
    \langle 1467\rangle },
\end{array}
\end{equation}
along with their cyclic and parity images ($7\times 14 = 98$ total quivers). In this case it suffices to take only these images since $i\to 8-i$ maps this set back to itself. And finally,
\begin{equation}
	\begin{array}{l}
 \frac{\langle 1236\rangle  \langle 1245\rangle }{\langle
   1234\rangle  \langle 1256\rangle }\to \frac{\langle
   1246\rangle  \langle 1345\rangle }{\langle 1234\rangle
    \langle 1456\rangle },
 \frac{\langle 1234\rangle  \langle 1256\rangle }{\langle
   1236\rangle  \langle 1245\rangle }\to \frac{\langle
   1235\rangle  \langle 1267\rangle  \langle 1456\rangle
   }{\langle 1236\rangle  \langle 1245\rangle  \langle
   1567\rangle }, \\
 \frac{\langle 1236\rangle  \langle 1245\rangle }{\langle
   1234\rangle  \langle 1256\rangle }\to \frac{\langle
   1246\rangle  \langle 2345\rangle }{\langle 1234\rangle
    \langle 2456\rangle },
 \frac{\langle 1234\rangle  \langle 1256\rangle }{\langle
   1236\rangle  \langle 1245\rangle }\to \frac{\langle
   1235\rangle  \langle 1267\rangle  \langle 2456\rangle
   }{\langle 1236\rangle  \langle 1245\rangle  \langle
   2567\rangle }, \\
 \frac{\langle 1234\rangle  \langle 1256\rangle }{\langle
   1236\rangle  \langle 1245\rangle }\to \frac{\langle
   1235\rangle  \langle 1267\rangle  \langle 3456\rangle
   }{\langle 1236\rangle  \langle 5(12)(34)(67)\rangle }
  ,
 \frac{\langle 1235\rangle  \langle 4567\rangle }{\langle
   5(12)(34)(67)\rangle }\to \frac{\langle 1236\rangle
   \langle 1245\rangle }{\langle 1234\rangle  \langle
   1256\rangle }, \\
 \frac{\langle 1235\rangle  \langle 1456\rangle }{\langle
   1256\rangle  \langle 1345\rangle }\to \frac{\langle
   1237\rangle  \langle 1245\rangle }{\langle 1234\rangle
    \langle 1257\rangle },
     \frac{\langle 1237\rangle  \langle 1245\rangle }{\langle
   1234\rangle  \langle 1257\rangle }\to \frac{\langle
   1247\rangle  \langle 2345\rangle }{\langle 1234\rangle
    \langle 2457\rangle }, \\
     \frac{\langle 1234\rangle  \langle 1257\rangle }{\langle
   1237\rangle  \langle 1245\rangle }\to \frac{\langle
   1235\rangle  \langle 1267\rangle  \langle 2457\rangle
   }{\langle 1237\rangle  \langle 1245\rangle  \langle
   2567\rangle },
 \frac{\langle 1236\rangle  \langle 1456\rangle }{\langle
   1256\rangle  \langle 1346\rangle }\to \frac{\langle
   1237\rangle  \langle 1246\rangle }{\langle 1234\rangle
    \langle 1267\rangle }, \\
 \frac{\langle 1234\rangle  \langle 1356\rangle }{\langle
   1236\rangle  \langle 1345\rangle }\to \frac{\langle
   1235\rangle  \langle 1367\rangle  \langle 3456\rangle
   }{\langle 1236\rangle  \langle 1345\rangle  \langle
   3567\rangle },
    \frac{\langle 1234\rangle  \langle 1356\rangle }{\langle
   1236\rangle  \langle 1345\rangle }\to \frac{\langle
   1235\rangle  \langle 1567\rangle  \langle 3456\rangle
   }{\langle 1256\rangle  \langle 1345\rangle  \langle
   3567\rangle }, \\
 \frac{\langle 1235\rangle  \langle 1567\rangle }{\langle
   1257\rangle  \langle 1356\rangle }\to \frac{\langle
   1237\rangle  \langle 1256\rangle  \langle 1345\rangle
   }{\langle 1234\rangle  \langle 1257\rangle  \langle
   1356\rangle },
 \frac{\langle 1234\rangle  \langle 1267\rangle  \langle
   1356\rangle }{\langle 1237\rangle  \langle 1256\rangle
    \langle 1346\rangle }\to \frac{\langle 1236\rangle
   \langle 1567\rangle  \langle 3456\rangle }{\langle
   1256\rangle  \langle 1346\rangle  \langle 3567\rangle
   },
\end{array}
\end{equation}
along with their dihedral and parity images ($14\times28 = 392$ total quivers).

The $A_2$ function evaluates to 504 distinct results on these 504 algebras, but
the 504 resulting quantities are not linearly independent:
there are 56 linear relationships amongst these $A_2$ functions.
It would be interesting to clarify the geometric origin of these
linear relations.
We conjecture, but lack the computer power to prove by explicit
computation, that these 504 quantities span the space of
nontrivial
weight-four
cluster functions for the $\Gr(4,7)$ algebra.

The $\Gr(4,7)$ algebras has 476 $A_3$ subalgebras~\cite{Golden:2013xva} on which we can evaluate $f_{A_3}$, but only 364 of these give distinct results.
We conjecture that these 364 quantities span the space of non-trivial weight-four cluster functions with
completely local coproducts having the desired Poisson structure properties (0 in $\Lambda^2 \B_2$ and $\pm 1$ in $\B_3 \otimes \mathbb{C}^*$).

We can list the 364 distinct $A_3$ evaluations by separating them in to three classes, and providing one (out of a possible six) generating quiver for each. First of all there are $14\times2 = 28$ $A_3$'s generated by the quivers
\begin{equation}
\begin{array}{l}
\frac{\langle 1236\rangle  \langle 1245\rangle }{\langle
   1234\rangle  \langle 1256\rangle }\to \frac{\langle 1234\rangle
   \langle 2456\rangle }{\langle 1246\rangle  \langle 2345\rangle
   }\to \frac{\langle 1256\rangle  \langle 2345\rangle  \langle
   4567\rangle }{\langle 1245\rangle  \langle 2567\rangle  \langle
   3456\rangle },\\ \frac{\langle 1256\rangle  \langle 2345\rangle  \langle 4567\rangle
   }{\langle 1245\rangle  \langle 2567\rangle  \langle 3456\rangle
   }\to \frac{\langle 1246\rangle  \langle 2567\rangle }{\langle
   1267\rangle  \langle 2456\rangle }\to \frac{\langle 1236\rangle
   \langle 1245\rangle }{\langle 1234\rangle  \langle 1256\rangle },
   \end{array}
 \end{equation}
along with their dihedral images.
Next there are $14\times6=84$ $A_3$'s generated by the quivers
\begin{equation}
\begin{array}{l}
 \frac{\langle 1245\rangle  \langle 3456\rangle }{\langle
   1456\rangle  \langle 2345\rangle }\to \frac{\langle 1235\rangle
   \langle 1456\rangle }{\langle 1256\rangle  \langle 1345\rangle
   }\to \frac{\langle 1234\rangle  \langle 1256\rangle }{\langle
   1236\rangle  \langle 1245\rangle }, \\
 \frac{\langle 1234\rangle  \langle 1257\rangle }{\langle
   1237\rangle  \langle 1245\rangle }\to \frac{\langle 1237\rangle
   \langle 1256\rangle }{\langle 1235\rangle  \langle 1267\rangle
   }\to \frac{\langle 1245\rangle  \langle 1567\rangle }{\langle
   1257\rangle  \langle 1456\rangle }, \\
 \frac{\langle 1267\rangle  \langle 1356\rangle }{\langle
   1236\rangle  \langle 1567\rangle }\to \frac{\langle 1346\rangle
   \langle 3567\rangle }{\langle 1367\rangle  \langle 3456\rangle
   }\to \frac{\langle 1236\rangle  \langle 1345\rangle }{\langle
   1234\rangle  \langle 1356\rangle }, \\
 \frac{\langle 1237\rangle  \langle 1245\rangle }{\langle
   1234\rangle  \langle 1257\rangle }\to \frac{\langle 1234\rangle
   \langle 2457\rangle }{\langle 1247\rangle  \langle 2345\rangle
   }\to \frac{\langle 1257\rangle  \langle 2345\rangle  \langle
   4567\rangle }{\langle 1245\rangle  \langle 2567\rangle  \langle
   3457\rangle },\\
 \frac{\langle 1237\rangle  \langle 1246\rangle }{\langle
   1234\rangle  \langle 1267\rangle }\to \frac{\langle 1234\rangle
   \langle 1267\rangle  \langle 1456\rangle }{\langle 1247\rangle
   \langle 1256\rangle  \langle 1346\rangle }\to \frac{\langle
   1256\rangle  \langle 1346\rangle  \langle 4567\rangle }{\langle
   1246\rangle  \langle 1567\rangle  \langle 3456\rangle }, \\
 \frac{\langle 1257\rangle  \langle 2345\rangle  \langle
   4567\rangle }{\langle 1245\rangle  \langle 2567\rangle  \langle
   3457\rangle }\to \frac{\langle 1247\rangle  \langle 2567\rangle
   }{\langle 1267\rangle  \langle 2457\rangle }\to \frac{\langle
   1237\rangle  \langle 1245\rangle }{\langle 1234\rangle  \langle
   1257\rangle },
\end{array}
\end{equation}
along with their cyclic and parity images.
Finally we have the $9\times28 = 252$ $A_3$'s generated by the quivers
\begin{equation}
	\begin{array}{l}
 \frac{\langle 1256\rangle  \langle 4567\rangle }{\langle
   1567\rangle  \langle 2456\rangle }\to \frac{\langle
   1246\rangle  \langle 1567\rangle }{\langle 1267\rangle
    \langle 1456\rangle }\to \frac{\langle 1236\rangle
   \langle 1245\rangle }{\langle 1234\rangle  \langle
   1256\rangle } ,
 \frac{\langle 1236\rangle  \langle 1245\rangle }{\langle
   1234\rangle  \langle 1256\rangle }\to \frac{\langle
   1234\rangle  \langle 1456\rangle }{\langle 1246\rangle
    \langle 1345\rangle }\to \frac{\langle 1256\rangle
   \langle 1345\rangle  \langle 4567\rangle }{\langle
   1245\rangle  \langle 1567\rangle  \langle 3456\rangle
   } ,\\
 \frac{\langle 1245\rangle  \langle 1567\rangle }{\langle
   1257\rangle  \langle 1456\rangle }\to \frac{\langle
   1235\rangle  \langle 1456\rangle }{\langle 1256\rangle
    \langle 1345\rangle }\to \frac{\langle 1234\rangle
   \langle 1257\rangle }{\langle 1237\rangle  \langle
   1245\rangle } ,
 \frac{\langle 1234\rangle  \langle 1257\rangle }{\langle
   1237\rangle  \langle 1245\rangle }\to \frac{\langle
   1237\rangle  \langle 1256\rangle }{\langle 1235\rangle
    \langle 1267\rangle }\to \frac{\langle 1245\rangle
   \langle 2567\rangle }{\langle 1257\rangle  \langle
   2456\rangle } ,\\
 \frac{\langle 1245\rangle  \langle 2567\rangle }{\langle
   1257\rangle  \langle 2456\rangle }\to \frac{\langle
   1235\rangle  \langle 2456\rangle }{\langle 1256\rangle
    \langle 2345\rangle }\to \frac{\langle 1234\rangle
   \langle 1257\rangle }{\langle 1237\rangle  \langle
   1245\rangle } ,
 \frac{\langle 1257\rangle  \langle 4567\rangle }{\langle
   1567\rangle  \langle 2457\rangle }\to \frac{\langle
   1247\rangle  \langle 1567\rangle }{\langle 1267\rangle
    \langle 1457\rangle }\to \frac{\langle 1237\rangle
   \langle 1245\rangle }{\langle 1234\rangle  \langle
   1257\rangle } ,\\
 \frac{\langle 1246\rangle  \langle 1567\rangle }{\langle
   1267\rangle  \langle 1456\rangle }\to \frac{\langle
   1236\rangle  \langle 1456\rangle }{\langle 1256\rangle
    \langle 1346\rangle }\to \frac{\langle 1234\rangle
   \langle 1267\rangle }{\langle 1237\rangle  \langle
   1246\rangle } ,
 \frac{\langle 1246\rangle  \langle 1567\rangle }{\langle
   1267\rangle  \langle 1456\rangle }\to \frac{\langle
   1236\rangle  \langle 2456\rangle }{\langle 1256\rangle
    \langle 2346\rangle }\to \frac{\langle 1234\rangle
   \langle 1267\rangle }{\langle 1237\rangle  \langle
   1246\rangle } ,\\
 \frac{\langle 1345\rangle  \langle 1567\rangle }{\langle
   1357\rangle  \langle 1456\rangle }\to \frac{\langle
   1235\rangle  \langle 3456\rangle }{\langle 1356\rangle
    \langle 2345\rangle }\to \frac{\langle 1234\rangle
   \langle 1357\rangle }{\langle 1237\rangle  \langle
   1345\rangle }
\end{array}
\end{equation}
along with their dihedral and parity images.

While this collection of functions is dramatically more tame than the vastly overcomplete space of completely general non-classical polylogarithms at weight 4, there are still 169 functional identities amongst these 364 $f_{A_3}$'s.
Again, it would be very interesting to understand these relations geometrically.

We now turn our attention to the two-loop 7-point MHV amplitude, whose coproduct was first calculated in \cite{Golden:2013xva}. The $B_3 \otimes \mathbb{C}^*$ portion of the coproduct can be separated into symmetric and antisymmetric parts under $\{x\}_3 \otimes y \to \{y\}_3 \otimes x$. The antisymmetric part, which corresponds to non-classical polylogarithms, can be fit to $A_3$ functions of the $\Gr(4,7)$ cluster algebra, and the symmetric part can be fit to $\Li_4(-\text{$\mathcal{X}$-coordinate})$.

The functional identities amongst $A_3$ functions prevent us from writing down a unique representation of $R_7^{(2)}$ at this point. We settle here for the shortest possible representation\footnote{Future developments
may reveal that a different, longer representation is ``better'' by manifesting other properties, either a physical property such as smooth behavior under the collinear limit~\cite{futureWork} or possibly even an additional, so far unnoticed mathematical property.}:
\begin{multline}\label{eq:7ptRep}
R_7^{(2)} \sim
	\frac{1}{2} f_{A_3}\left(\textstyle{\frac{\langle 1245\rangle  \langle
   1567\rangle }{\langle 1257\rangle  \langle 1456\rangle }},
   \frac{\langle 1235\rangle  \langle 1456\rangle }{\langle
   1256\rangle  \langle 1345\rangle }, \frac{\langle 1234\rangle
   \langle 1257\rangle }{\langle 1237\rangle  \langle 1245\rangle
   }\right)+\frac{1}{2} f_{A_3}\left(\textstyle{\frac{\langle 1345\rangle  \langle
   1567\rangle }{\langle 1357\rangle  \langle 1456\rangle }},
   \frac{\langle 1235\rangle  \langle 3456\rangle }{\langle
   1356\rangle  \langle 2345\rangle }, \frac{\langle 1234\rangle
   \langle 1357\rangle }{\langle 1237\rangle  \langle 1345\rangle
   }\right)\\
   +\text{Li}_4\left(-\textstyle{\frac{\langle 1234\rangle  \langle 1256\rangle
   }{\langle 1236\rangle  \langle 1245\rangle
   }}\right)+\text{Li}_4\left(-\textstyle{\frac{\langle 1234\rangle  \langle
   1257\rangle }{\langle 1237\rangle  \langle 1245\rangle
   }}\right)+\frac{1}{4}
   \text{Li}_4\left(-\textstyle{\frac{\langle 1234\rangle  \langle 1357\rangle
   }{\langle 1237\rangle  \langle 1345\rangle }}\right)+\frac{1}{4}
   \text{Li}_4\left(-\textstyle{\frac{\langle 1234\rangle  \langle 1456\rangle
   }{\langle 1246\rangle  \langle 1345\rangle }}\right)\\
   -\frac{1}{4} \text{Li}_4\left(-\textstyle{\frac{\langle 1234\rangle
   \langle 1257\rangle  \langle 1356\rangle }{\langle 1237\rangle
   \langle 1256\rangle  \langle 1345\rangle }}\right)+\frac{1}{4}
   \text{Li}_4\left(-\textstyle{\frac{\langle 1234\rangle  \langle 1267\rangle
   \langle 1356\rangle }{\langle 1237\rangle  \langle 1256\rangle
   \langle 1346\rangle }}\right)+\frac{1}{4}
   \text{Li}_4\left(-\textstyle{\frac{\langle 1235\rangle  \langle 1456\rangle
   }{\langle 1256\rangle  \langle 1345\rangle }}\right)\\-\frac{1}{4}
   \text{Li}_4\left(-\textstyle{\frac{\langle 1234\rangle  \langle 1257\rangle
   \langle 1456\rangle }{\langle 1247\rangle  \langle 1256\rangle
   \langle 1345\rangle }}\right)-\frac{1}{4}
   \text{Li}_4\left(-\textstyle{\frac{\langle 1234\rangle  \langle 1267\rangle
   \langle 1456\rangle }{\langle 1247\rangle  \langle 1256\rangle
   \langle 1346\rangle }}\right)-\frac{1}{4}
   \text{Li}_4\left(-\textstyle{\frac{\langle 1234\rangle  \langle 1457\rangle
   }{\langle 1247\rangle  \langle 1345\rangle }}\right)\\ +\text{ dihedral} +\text{parity~conjugate.}
\end{multline}
As indicated by the $\sim$, this result expresses the ``most complicated part'' of $R_7^{(2)}$---the
difference between the function presented here and the actual amplitude is some weight-four
polynomial in the functions $-\Li_k(-x)$ for $k=1,2,3$ (and $\pi^2$), with arguments $x$ drawn from the
385 $\mathcal{X}$-coordinates of the $\Gr(4,7)$ cluster algebra.

We end this section by reiterating some important features of the $A_2$ and $A_3$ functions as conveyed in eq.~(\ref{eq:7ptRep}). Given the known symbol of $R_7^{(2)}$ there is no difficulty in principle to find a representation of the non-classical component of this amplitude in terms of (for example) the collection $\Li_{2,2}$ functions (see the appendix) with simple ratios of $\mathcal{X}$-coordinates as arguments. The problem with fitting the non-classical portion of the amplitude to some general basis of this type is that these functions in general have non-$\mathcal{A}$ coordinates as entries in their symbols.  Therefore, the remaining classical $\Li$'s needed to express the full amplitude could then have arbitrarily complicated algebraic functions of $\mathcal{X}$-coordinates as arguments, which makes constructing an ansatz exceptionally difficult. The $A_2$ function solves this problem because it has only $\mathcal{A}$-coordinates in its symbol, therefore providing a basis which is sufficient to capture the non-classical component while ensuring that the remaining classical $\Li$'s can be taken to have only (minus) $\mathcal{X}$-coordinates as arguments. The packaging of $A_2$ functions into $A_3$'s manifests even more structure of the amplitude $R_7^{(2)}$---namely the complete (i.e., term-by-term) locality and Poisson structure of its coproduct components.

\section{Conclusion}

Motivated by the cluster structure apparently underlying the structure of amplitudes in
SYM theory~\cite{Golden:2013xva},
in this paper we defined and studied the simplest few examples of \emph{cluster polylogarithm functions}
at transcendentality weight four.  We found that the $A_2$ algebra admits
a single non-trivial function $f_{A_2}$ of this type, and for several other cluster algebras
which we were able to analyze by explicit computation we found that the space of cluster functions
is spanned by $f_{A_2}$ evaluated on all available $A_2$ subalgebras. Interestingly, we found that these functions all have ``Stasheff polytope local'' $B_3 \otimes \mathbb{C}^*$ content which can be expressed in
terms of $\{x\}_3 \otimes y - \{y\}_3 \otimes x$ with pairs $x,y$ having Poisson
bracket 1 (and therefore associated to pentagonal faces of the appropriate generalized
Stasheff polytope).

We then considered an even more special collection of ``Stasheff polytope local'' functions which have
$\Lambda^2 \B_2$ content expressible in terms of $\{x\}_2 \wedge \{y\}_2$ with $x,y$ having
Poisson bracket 0 (and therefore associated to quadrilateral faces).
For the $A_3$ algebra we found a unique nontrivial function $f_{A_3}$ with this property, and conjectured
that the space of such functions for more general algebras is spanned by
the function $f_{A_3}$ evaluated
on all available $A_3$ subalgebras.

Obviously it would be of mathematical interest to further explore these classes of functions, as well
as suitable generalizations of them at higher weight and for more general cluster algebras (especially
algebras of infinite type).

We used the $A_3$ function to write an explicit formula for the ``most complicated part''
of the two-loop 7-particle MHV amplitude in SYM theory.   We are confident that the $A_3$ function
suffices to similarly express two-loop MHV amplitudes for all $n$, both because we have
checked some cases explicitly but more importantly because we know~\cite{futureWork,SimonsTalk}
that these amplitudes
have completely local coproducts in the sense mentioned a moment ago.

However a number of important questions about the cluster structure of these amplitudes remain.  For example, attention was called in~\cite{Golden:2013xva} to the curious fact that the $\Lambda^2 B_2$ component of the 7-particle amplitude can be written as a 42-term linear combination of $\{x_i\}_2 \wedge \{x_j\}_2$ involving only 42 out of the 1785 distinct pairs of Poisson commuting $\mathcal{X}$-coordinates available in the $\Gr(4,7)$ cluster algebra. It was natural to wonder whether there is any characteristic of these 42 which distinguishes them from the rest, and which might be able to explain ``why'' the amplitude's coproduct can be expressed in terms of only these 42.  Unfortunately we are no closer to answering this question than~\cite{Golden:2013xva} was.  The first obstacle is that the formula (5.2) in~\cite{Golden:2013xva} is not manifestly expressible in terms of $A_3$ functions:  there does not exist any $A_3$ subalgebra of $\Gr(4,7)$ which has a quadrilateral face corresponding to the first term in (5.2) (nor to any of its symmetric images).  In contrast, if we evaluate the $\Lambda^2 B_2$ content of the amplitude by starting with eq.~(\ref{eq:7ptRep}) and associating to each $f_{A_3}$ the corresponding coproduct component shown in eq.~(\ref{eq:A3motivic}) we obtain a 56-term linear combination which is nontrivially equal to the 42-term expression presented in (5.2) of~\cite{Golden:2013xva}\footnote{It turns out that 14 terms in the former correspond exactly to the 14 terms on the second line of (5.2) of the latter; so perhaps it would be better to say that the nontrivial relation is between a 42-term expression and a 28-term expression in $\Lambda^2 B_2$.}.  Also, the results of this paper unfortunately shed no light on the curiosity noted in~\cite{Golden:2013xva} that for both $n=6,7$, the coproduct of the $n$-particle two-loop MHV amplitude can be expressed in terms of only 3/5 of the $\mathcal{X}$-coordinates available in the $\Gr(4,n)$ cluster algebra.

Our exploration of the appropriate function space for two-loop MHV amplitudes at arbitrary $n$ was
strongly motivated by a similar exploration of functions appropriate for non-MHV and higher-loop $n=6$
amplitudes by Dixon and collaborators~\cite{Dixon:2011pw,Dixon:2011nj,Dixon:2013eka}.
It would be very interesting to explore the (necessarily very close) relationship between
their ``hexagon functions'' and the various cluster functions we have explored, which we leave
to future work.
Here however we have focused exclusively on purely mathematical
constraints:  the $\mathcal{A}$-coordinate condition on symbols, the $\mathcal{X}$-coordinate
condition on functions, and the locality and Poisson structure constraints on the coproduct.
These are listed in order of increasing mathematical power, but also in order of increasing
physical obscurity.  We confess to having no physical explanation
of why SYM theory should select weight-four
polylogarithm functions whose coproducts are local in the
generalized Stasheff polytope or have any particular relation to the Poisson structure, except to speculate that it might be related to the integrability of SYM theory.  Notice also that clusters represent sets of coordinates that are compatible in some way. For instance, it is known~\cite{FG03b} that for $\Gr(2,n)$ the cluster structure is isomorphic to that of polygon triangulations, and that in turn to planar tree diagrams. To each tree corresponds a cluster, which can therefore be thought of as a channel for the tree amplitude. Cluster coordinates are then compatible in the sense that they correspond to possible simultaneous poles in planar scattering. Perhaps some more sophisticated version of this argument will hold here. It is natural to wonder if there exists an alternative formulation for SYM theory
amplitudes which makes these (and perhaps other, still hidden) cluster algebraic properties manifest.

With our current understanding of how to write down the most
complicated part of the two-loop MHV amplitudes it is reasonable to contemplate finding
fully analytic expressions for them.
To this end the next step
is to begin applying various physical constraints to fix ambiguities involving products
of functions of lower weight as well as beyond-the-symbol terms.
The most obvious such constraints include the first- and last-entry conditions on the
symbol,
the requirement of smooth behavior under collinear limits,
and especially
the highly constraining requirement of analyticity inside the Euclidean kinematic
region.
We believe the last of these, in particular, might be strong enough to
fix a unique (or almost unique) ``analytic tail'' to the $A_3$ function, perhaps similar
in form to the analytic tail which appears in the $L(x^+,x^-)$ building block of the
GSVV formula~\cite{Goncharov:2010jf}. Adding these terms of lower weight will help us resolve the ambiguities present in eq.~(\ref{eq:7ptRep}), where we had to arbitrarily choose one out of many possible representations in terms of $A_3$ functions.   Moreover
we suspect that ``the right'' completion of the $A_3$ function (once it is found) will
continue to be the unique non-classical
building block for all $n$-particle two-loop MHV amplitudes.

Based on the surprising fact that the fundamental building block of the two-loop MHV amplitudes
seems to be a function involving only $n=6$ particles, it is natural to hope that the
available results on higher-loop and NMHV functions for $n=6$,
when supplemented by suitable ``cluster algebraic'' constraints of the type we have discussed in this paper, may serve as a springboard
for unlocking the structure of $n$-particle MHV and NMHV amplitudes at higher loop order.

\acknowledgments

We have benefitted from stimulating discussions with Nima Arkani-Hamed,
Lance Dixon, James Drummond, Cristian Vergu, and especially Alexander Goncharov.
MS and AV are grateful to the CERN theory group for hospitality during the initial
stages of this work.
This work was supported by the US Department of Energy under contracts
DE-SC0010010 Task A (JG, MS) and DE-FG02-11ER41742 Early Career Award (MP, AV),
and by the
Sloan Research Foundation (AV).

\appendix

\section{Functional representatives}

We present here functional representations for the $A_2$ and $A_3$ functions
studied in the paper. These functions are completely defined (as always, modulo products of lower-weight
functions) by their coproducts, shown in eqs.~(\ref{eq:pentagonmotivic})
and~(\ref{eq:A3motivic}), but some readers may be comforted by seeing concrete
functional representations for them.
However, we relegate these formulas to the appendix because they are provided as is,
with no express or implied warranty,
and
certainly not the implied warranty of suitability for numerically evaluating actual
SYM theory amplitudes.  For such an application one would first need to append to each of the
functions shown below a suitable ``analytic tail'' comprising a carefully chosen product
of lower-weight functions, specially crafted to give the functions the right analytic properties.
Nevertheless we do believe that these functions capture the ``most complicated part'' of all
two-loop MHV amplitudes in SYM theory.

There are several different types of generalized polylogarithm functions in terms of
which non-classical functions can be expressed.
At weight 4 it suffices to use
the function $\Li_{2,2}(x,y)$
(see for example~\cite{Duhr:2011zq} for a discussion), whose symbol is
\begin{align}
        &(y-1)\otimes (x-1)\otimes x\otimes y+(y-1)\otimes (x-1)\otimes y\otimes x+(y-1)\otimes y\otimes (x-1)\otimes x\nonumber\\
        &-(x y-1)\otimes (x-1)\otimes x\otimes y-(x y-1)\otimes (x-1)\otimes y\otimes x-(x y-1)\otimes x\otimes (x-1)\otimes x\nonumber\\
        &+(x y-1)\otimes x\otimes x\otimes x+(x y-1)\otimes x\otimes x\otimes y+(x y-1)\otimes x\otimes (y-1)\otimes y\nonumber\\
        &+(x y-1)\otimes x\otimes y\otimes x+(x y-1)\otimes (y-1)\otimes x\otimes y+(x y-1)\otimes (y-1)\otimes y\otimes x\nonumber\\
        &-(x y-1)\otimes y\otimes (x-1)\otimes x+(x y-1)\otimes y\otimes x\otimes x+(x y-1)\otimes y\otimes (y-1)\otimes y.
\end{align}

\subsection{The $A_2$ function}

The $A_2$ function may be represented as
\begin{align}
\label{eq:pentagonfunction}
f_{A_2}\sim\sum_{i,j}^5 j L_{2,2}(x_i,x_{i+j})
\end{align}
in terms of
\begin{multline}
\label{eq:weird}
        L_{2,2}(x,y)=\frac{1}{2} \text{Li}_{2,2}\left(\frac{x}{y},-y\right)+\frac{1}{6} \left(\text{Li}_4\left(\frac{1+x}{x y}\right)+\text{Li}_4\left(\frac{x (1+y)}{y(1+x)}\right)\right) \\
+\frac{1}{5} \left(\text{Li}_4\left(\frac{1+x}{x y}\right)+\frac{1}{2} \text{Li}_4\left(\frac{1+x}{1+y}\right)\right) +\frac{1}{2}\text{Li}_3\left(\frac{x}{y}\right)\log\left(\frac{1+x}{1+y}\right)-(x \leftrightarrow y).
\end{multline}
The factor of $j$ in the summand may seem awkward, but
when fully expanded out the sum generates a total of 20 $\Li_{2,2}$ terms, each
with coefficient
$\pm \frac{3}{2}$ or $\pm \frac{1}{2}$ (each possibility occurs five times).
Note that the function $L_{2,2}$ has the simple
coproduct $\delta L_{2,2}(x,y)\rvert_{\Lambda^2 \B_2} = \{x\}_2 \wedge \{y\}_2$ (it is therefore very similar to Goncharov's $\kappa(x,y)$ function \cite{Goncharov-Galois1}).
The rather strange looking $\Li_4$ terms in eq.~(\ref{eq:weird}) of course make no contribution
to $\Lambda^2 \B_2$; they are carefully tuned to ensure that eq.~(\ref{eq:pentagonfunction})
has clustery $\B_3 \otimes \mathbb{C}^*$ content. The $\Li_3 \cdot \log$ terms are of course irrelevant
inside $\mathcal{L}_4$, but they are required for $f_{A_2}$ to be a cluster $\mathcal{A}$-function of the $A_2$ algebra.
The symbol of eq.~(\ref{eq:pentagonfunction}) is not identical to the one shown in eq.~(\ref{eq:A2symbol}), but the difference
between the two is
annihilated by $\rho$ (i.e., they differ by products of functions of lower weight).

\subsection{The $A_3$ function}

The $A_3$ function may of course be written as the sum of eq.~(\ref{eq:weird})'s for the six pentagons in $A_3$, but the simple form of $\delta f_{A_3}(x_1,x_2,x_3) \rvert_{\Lambda^2 \B_2}$ suggests that there is a more concise functional representation. Indeed, we find that a representative of the $A_3$ function can be written as
\begin{equation}\label{eq:fA3func}
        f_{A_3}(x_1,x_2,x_3)\sim\sum_{i=1}^3 K_{2,2}(x_{i,1},x_{i,2}) + \frac{1}{2}\sum_{i=1}^6 (-1)^i \Li_4(-e_i)
\end{equation}
where the $x_{i,j}$ and $e_i$ are defined in~(\ref{eq:A3coords}) and we use here the new combination
\begin{equation}
K_{2,2}(x,y) = \frac{1}{2}\Li_{2,2} (x/y,-y)- \Li_4(x/y) -\frac{2}{3}\Li_3(x/y)\log(y)-(x\leftrightarrow y).
\end{equation}
As was the case for the $A_2$ function, the $\Li_3 \cdot \log$ terms are chosen so that the symbol of~(\ref{eq:fA3func}) is expressible entirely in terms of cluster $\mathcal{A}$-coordinates of the $A_3$ algebra.

\end{document}